\pdfoutput=1

\documentclass[letterpaper,twocolumn,10pt]{article}
\usepackage{usenix,epsfig,endnotes}

\usepackage{cite}
\usepackage{graphicx}
\usepackage{float}
\usepackage{subcaption}
\usepackage{bmpsize}
\usepackage[textsize=tiny,obeyDraft]{todonotes}
\usepackage{amsmath}
\usepackage{multirow}

\usepackage{algorithm}
\usepackage{algpseudocode}

\usepackage{url}
\usepackage{cleveref}

\newcommand{\DownsampleFactor}{10}

\begin{document}

%don't want date printed
\date{}

%
% paper title
% can use linebreaks \\ within to get better formatting as desired
% Do not put math or special symbols in the title.
\title{PowerSpy: Location Tracking using Mobile Device Power Analysis}
% another possible title:
% PowerSpy: Location and routes inferred from mobile device power consumption

% author names and affiliations
% use a multiple column layout for up to three different
% affiliations
% \author{}
\author{
  {\rm Yan Michalevsky, Aaron Schulman,} \\
   {\rm Gunaa Arumugam Veerapandian and Dan Boneh} \\
    \small Computer Science Department\\
    \small Stanford University
    \and
    {\rm Gabi Nakibly}\\
    \small National Research and Simulation Center\\
    \small Rafael Ltd.
} % end author

% conference papers do not typically use \thanks and this command
% is locked out in conference mode. If really needed, such as for
% the acknowledgment of grants, issue a \IEEEoverridecommandlockouts
% after \documentclass

% for over three affiliations, or if they all won't fit within the width
% of the page, use this alternative format:
%
%\author{\IEEEauthorblockN{Michael Shell\IEEEauthorrefmark{1},
%Homer Simpson\IEEEauthorrefmark{2},
%James Kirk\IEEEauthorrefmark{3},
%Montgomery Scott\IEEEauthorrefmark{3} and
%Eldon Tyrell\IEEEauthorrefmark{4}}
%\IEEEauthorblockA{\IEEEauthorrefmark{1}School of Electrical and Computer Engineering\\
%Georgia Institute of Technology,
%Atlanta, Georgia 30332--0250\\ Email: see http://www.michaelshell.org/contact.html}
%\IEEEauthorblockA{\IEEEauthorrefmark{2}Twentieth Century Fox, Springfield, USA\\
%Email: homer@thesimpsons.com}
%\IEEEauthorblockA{\IEEEauthorrefmark{3}Starfleet Academy, San Francisco, California 96678-2391\\
%Telephone: (800) 555--1212, Fax: (888) 555--1212}
%\IEEEauthorblockA{\IEEEauthorrefmark{4}Tyrell Inc., 123 Replicant Street, Los Angeles, California 90210--4321}}

% use for special paper notices
%\IEEEspecialpapernotice{(Invited Paper)}

% make the title area
\maketitle

% As a general rule, do not put math, special symbols or citations
% in the abstract
\subsection*{Abstract}
Modern mobile platforms like Android enable applications to read
aggregate power usage on the phone.  This information is considered
harmless and reading it requires no user permission or notification.
We show that by simply reading the phone's aggregate power consumption
over a period of a few minutes an application can learn information
about the user's location.  Aggregate phone power consumption data is
extremely noisy due to the multitude of components and applications
that simultaneously consume power.  Nevertheless, by using machine
learning algorithms we are able to successfully infer the phone's
location. We discuss several ways in which this privacy leak can be
remedied.

% no keywords

% For peer review papers, you can put extra information on the cover
% page as needed:
% \ifCLASSOPTIONpeerreview
% \begin{center} \bfseries EDICS Category: 3-BBND \end{center}
% \fi
%
% For peerreview papers, this IEEEtran command inserts a page break and
% creates the second title. It will be ignored for other modes.
% \IEEEpeerreviewmaketitle

\section{Introduction}
% no \IEEEPARstart

Our phones are always within reach and their location is mostly
the same as our location.  In effect, tracking the location of a
phone is practically the same as tracking the location of its
owner.  Since users generally prefer that their location not be
tracked by arbitrary 3rd parties, all mobile platforms consider the
device's location as sensitive information and go to considerable
lengths to protect it: applications need explicit user permission to
access the phone's GPS and even reading coarse location data based on
cellular and WiFi connectivity requires explicit user permission.

In this work we show that despite these restrictions applications can
covertly learn the phone's location.  They can do so using a seemingly
benign sensor: the phone's power meter that measures the phone's power
consumption over a period of time.  Our work is based on the
observation that the phone's location significantly affects the power
consumed by the phone's cellular radio. The power consumption is
affected both by the distance to the cellular base station to which
the phone is currently attached (free-space path loss) and by
obstacles, such as buildings and trees, between them (shadowing).  The
closer the phone is to the base station and the fewer obstacles
between them the less power the phone consumes.  The strength of the
cellular signal is a major factor affecting the power used by the
cellular radio~\cite{Schulman2010}.  Moreover, the cellular radio is
one of the most dominant power consumers on the
phone~\cite{huang-mobisys12}.

Suppose an attacker measures in advance the power profile consumed by
a phone as it moves along a set of known routes or in a predetermined
area such as a city.  We show that this enables the attacker to infer
the target phone's location over those routes or areas by simply
analyzing the target phone's power consumption over a period of time.
This can be done with no knowledge of the base stations to which the
phone is attached. 
% as long as the attacker knows the general area
% in which the victim moves.

A major technical challenge is that power is consumed simultaneously
by many components and applications on the phone in addition to the
cellular radio.  A user may launch applications, listen to music, turn
the screen on and off, receive a phone call, and so on.  All these
activities affect the phone's power consumption and result in a very
noisy approximation of the cellular radio's power usage.  Moreover,
the cellular radio's power consumption itself depends on the phone's
activity, as well as the distance to the base-station: during a voice
call or data transmission the cellular radio consumes more power than
when it is idle. All of these factors contribute to the phone's power
consumption variability and add noise to the attacker's view: the
power meter only provides aggregate power usage and cannot be used to
measure the power used by an individual component such as the cellular radio.

Nevertheless, using machine learning, we show that the
phone's aggregate power consumption over time completely reveals the
phone's location and movement.  Intuitively, the reason why all this
noise does not mislead our algorithms is that the noise is not
correlated with the phone's location.  Therefore, a sufficiently long
power measurement (several minutes) enables the learning algorithm to
``see'' through the noise.
We refer to power consumption measurements as time-series and use methods
for comparing time-series to obtain classification and pattern matching algorithms
for power consumption profiles.

In this work we use machine learning to identify the
routes taken by the victim based on previously collected power
consumption data.  We study three types of user tracking goals:
\begin{enumerate}
\item \textbf{Route distinguishability:} First, we ask whether
an attacker can tell what route the user is taking among a fixed
set of possible routes.

\item \textbf{Real-time motion tracking:} Assuming the user is taking
  a certain known route, we ask whether an attacker can identify her
  location along the route and track the device's position on the
  route in real-time.

\item \textbf{New route inference:} Finally, suppose a user is moving
  along an arbitrary (long) route.  We ask if an attacker can learn
  the user's route using the previously measured power profile of many
  (short) road segments in that area.  The attacker 
  composes the power profile of the short road segments to identify
  the user's route and location at the end of the route.
\end{enumerate}

\noindent
We emphasize that our approach is based on measuring the phone's
aggregate power consumption and nothing else.  In particular, we do not use the
phone's signal strength as this data is protected on Android and
iOS devices and reading it requires user permission.
In contrast, reading the phone's power meter requires no special permissions.  

On Android reading the phone's aggregate power
meter is done by repeatedly reading the following two files: \\
\mbox{}\quad {\tt\small
  /sys/class/power\_supply/battery/voltage\_now \\
\mbox{}\quad
  /sys/class/power\_supply/battery/current\_now} \\
Over a hundred applications in the Play Store access these files.
While most of these simply monitor battery usage, our work shows that
all of them can also easily track the user's location.

\medskip\noindent
{\bf Our contributions.} Our work makes the following contributions:
\begin{itemize}
\item We show that the power meter available on modern phones can
  reveal potentially private information.
  % Up until now, external
  % power measurements have mostly been used to attack
  % cryptosystems~\cite{kocher1999differential}.

\item We develop the machine learning techniques needed to use data
  collected from the power meter to infer location information.  The
  technical details of our algorithms are presented in
  sections~\ref{sec:route-distinguishability},
  \ref{sec:mobile-device-tracking} and~\ref{sec:newroutes}, followed
  by experimental results.

\item In sections~\ref{sec:future} and~\ref{sec:defenses} we discuss
  potential continuation to this work, as well as defenses to prevent
  this type of information leakage.
\end{itemize}

% \medskip
% The rest of the paper is organized as follows: We start with defining
% the threat model.  Then we provide technical background about signal
% strength and power consumption, and relate it to our method.  We
% follow with stating the underlying assumptions behind our research.

\section{Threat Models}

We assume a malicious application is installed on the victim's device
and runs in the background.  The application has no permission to
access the GPS or any other location data such as the cellular or WiFi
components. In particular, the application has no permission to query
the identity of visible cellular base stations or the SSID of
visible WiFi networks.

We only assume access to power data (which requires no special
permissions on Android) and permission to communicate with a remote
server.  Network connectivity is needed to generate dummy low rate
traffic to prevent the cellular radio from going into low power state.
In our setup we also use network connectivity to send
data to a central server for processing.  However, it may be
possible to do all processing on the phone.\footnote{It is important to mention
here that while a network access permission will appear in the permission
list for an installed application, it does not currently appear in the list
of required permissions prior to application installation.}

As noted earlier, the application can only read the {\em aggregate}
power consumed by the phone.  It cannot measure the power consumed by the
cellular radio alone.   This presents a significant challenge since many
components on the phone consume variable amounts of power at any given time.
Consequently, all the measurements are extremely noisy and we need a way
to ``see'' though the noise.

% At the time of this writing, 179 applications submitted to
% the Google Play store read the voltage and current data. We assume
% the majority of them simply profile application power use to prolong
% battery life.

To locate the phone, we assume the attacker has prior knowledge of the
area or routes through which the victim is traveling. This knowledge
allows the attacker to measure the power consumption profile of
different routes in that area in advance. Our system correlates this
data with the phone's measured power usage and we show that, despite
the noisy measurements, we are able to correctly locate the phone.
Alternatively, as for many other machine learning cases, the training 
data can also be collected after obtaining the unlabeled query data.
For instance, an attacker obtained a power consumption profile of a
user, the past location of whom it is extremely important to determine.
She can still collect, after the fact, reference profiles for a limited area in which the user
has likely been driving and carry out the attack.

For this to work we need the tracked phone to be moving by a car or a bus
while being tracked.  Our system cannot locate a phone that is
standing still since that only provides the power profile for a single
location. We need multiple adjacent locations for the attack to work.

Given the resources at our disposal, the focus of this work is on
locating a phone among a set of local routes in a
pre-determined area.  A larger effort is needed to scale the system to
cover the entire world by pre-measuring the power profile of all road
segments worldwide.  Nevertheless, our localized experiments already
show that tracking users who follow a daily routine is quite possible.
For example, a mobile device owner might choose one of a small number
of routes to get from home to work. The system correctly identifies
what route was chosen and in real-time identifies where the phone
is along that route.  This already serves as a cautionary note about
the type of information that can be leaked by a seemingly innocuous sensor
like the power meter.

We note that scaling the system to cover worldwide road segments can
be done by crowd-sourcing: a popular app, or perhaps even the core OS,
can record the power profile of streets traveled by different users
and report the results to a central server.  Over time the resulting
dataset will cover a significant fraction of the world.  On the
positive side, our work shows that service providers can legitimately
use this dataset to improve the accuracy of location services.  On
the negative side, tracking apps can use it to covertly locate users.
Given that all that is required is one widespread application, many
actors in the mobile space are in a position to build the required
dataset of power profiles and use it as they will.

% A less restrictive threat model is one in which the malicious application is capable of identifying the routes
% part of the time (through occasional access to GPS or WiFi access point identifiers),
% and thus, is capable of learning the power consumption profiles for those routes directly through the user's
% own device. That might be a simpler case, since power profiles measured for the same route on the same device
% are likely to exhibit even greater similarity.\todo{Are we addressing both threat models? One of them?}

\section{Background}

In this section we provide technical background on the relation between a
phone's location and its cellular power consumption. We start with a
description of how location is related to signal strength, then we
describe how signal strength is related to power consumption.
Finally, we present examples of this phenomenon, and we demonstrate how
obtaining access to power measurements could leak information about a
phone's location.

\subsection{Location affects signal strength and power consumption}
Distance to the base station is the primary factor that determines a phone's signal strength.
The reason for this is, for signals propagating in free space, the signal's power loss is proportional to the square of the distance it travels over~\cite{goldsmith2005wireless}.
%and is given (in the simplest form) by
%$$ \frac{P_r}{P_t} \propto \frac{G_l \cdot \lambda^2}{(4 \pi d)^2} $$
%where $P_r, P_t$ are the recieved and transmitted power respectively, $G_l$ is related to the transmit
%and receive antenna properties, and $d$ is the distance of the receiver from the transmitter.
Signal strength is not only determined by path loss, it is also affected by objects in the signal path, such as trees and buildings, that attenuate the signal. Finally, signal strength also depends on multi-path interference caused by objects that reflect the radio signal back to the phone through various paths having different lengths.

In wireless communication theory signal strength is often modeled as random variation (e.g., log-normal shadowing~\cite{goldsmith2005wireless})
to simulate many different environments\footnote{Parameters of the model can be calibrated to better match a specific environment of interest.}.
However, in one location signal strength can be fairly consistent as base stations, attenuators, and reflectors are mostly stationary.

A phone's received signal strength to its base station affects its cellular modem power consumption.
Namely, phone cellular modems consume less instantaneous power when transmitting and receiving at high signal strength compared to low signal strength.
Schulman et. al.~\cite{Schulman2010} observed this phenomenon on several different cellular devices operating on different cellular protocols.
They showed that communication at a poor signal location can result in a device power draw that
is 50\% higher than at a good signal location.

The primary reason for this phenomenon is the phone's power amplifier used for transmission which increases its gain as signal strength drops~\cite{goldsmith2005wireless}. This effect also occurs when a phone is only receiving packets. The reason for this is cellular protocols which require constant transmission of channel quality and acknowledgments to base stations.
%There also is increased power consumption due to increased processing that is put toward error correcing low signal strength signals.

\subsection{Power usage can reveal location}

The following results from driving experiments demonstrate the potential of leaking location from power measurements.

\begin{figure*}
	\centering
%  \begin{subfigure}{0.4\textwidth}
    \includegraphics[width=0.4\textwidth,height=120px]{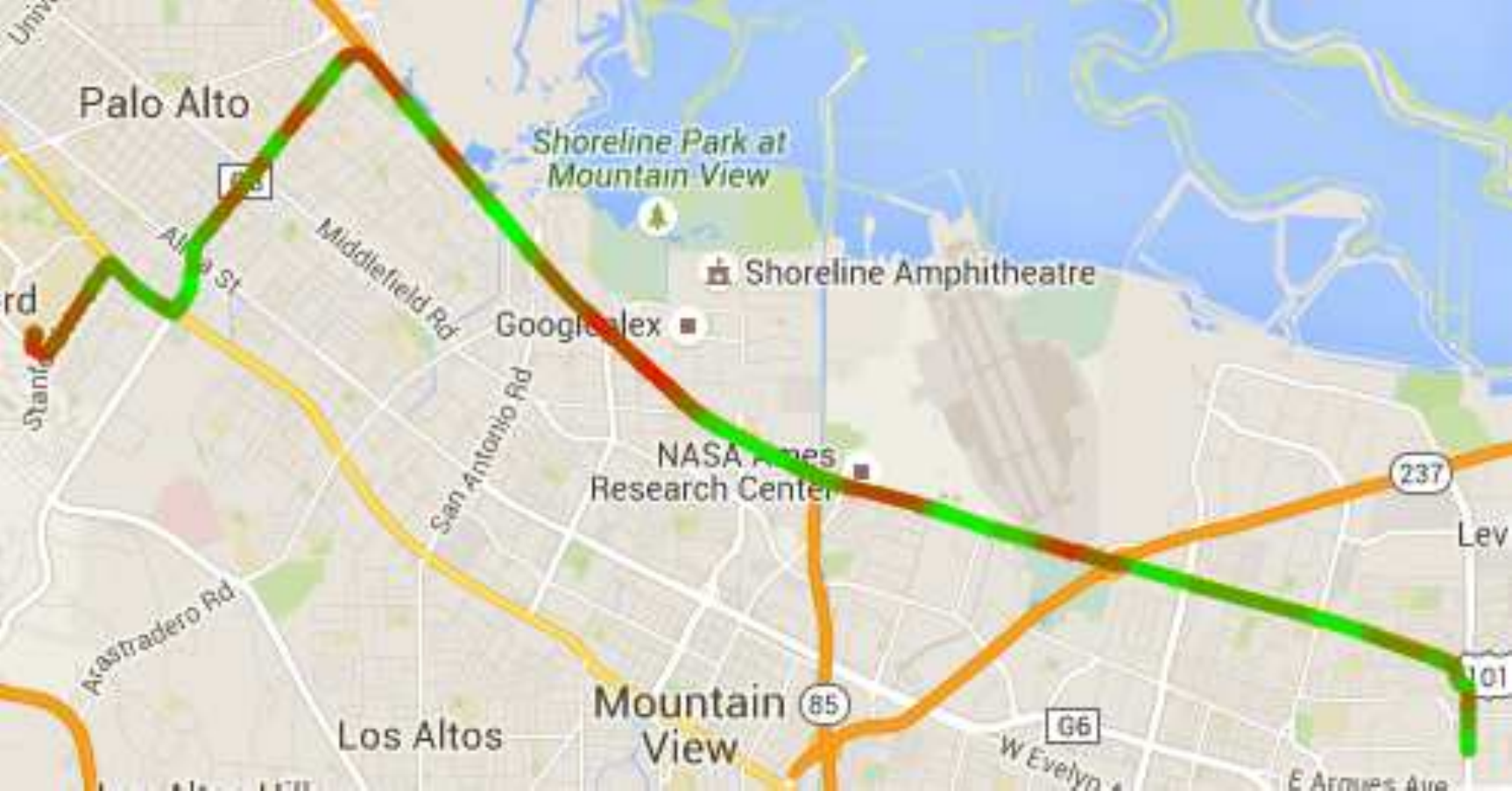}
%  \end{subfigure}
  \hspace{0.5cm}
%  \begin{subfigure}{0.4\textwidth}
    \includegraphics[width=0.4\textwidth,height=120px]{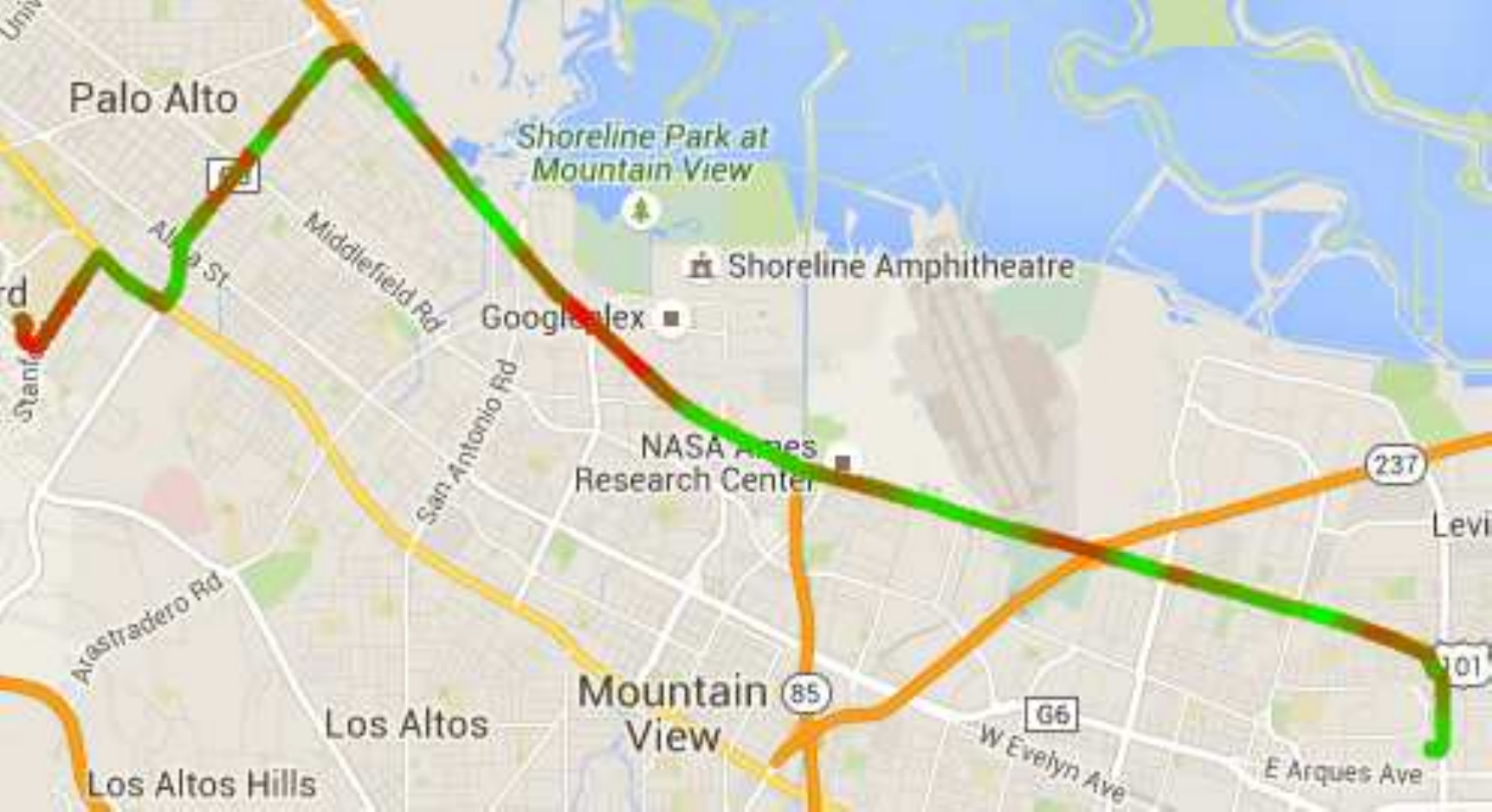}
%  \end{subfigure}
  \caption{Signal strength profiles measured on two different days are stable.}
  \label{fig:signal-stability}
\end{figure*}

We first demonstrate that signal strength in each location on a drive can be
static over the course of several days. We collected signal strength measurements from a smartphone once, and
again several days later. In Figure~\ref{fig:signal-stability} we plot the signal
strength observed on these two drives. In this figure it is apparent that (1)
the segments of the drive where signal strength is high (green) and low (red)
are in the same locations across both days, and (2) that the progression of
signal strength along the drive appears to be a unique irregular pattern.

Next, we demonstrate that just like signal strength, power measurements of a
smartphone, while it communicates, can reveal a stable, unique pattern for a
particular drive. Unlike signal strength, power measurements are less likely to be
stable across drives because power depends on how the cellular modem reacts to
changing signal strength: a small difference in signal strength between two
drives may put the cellular modem in a mode that has a large difference in
power consumption. For example, a small difference in signal strength may cause a phone to hand-off to a 
different cellular base station and stay attached to it for some time (Section~\ref{sec:hysterisis}).

\begin{figure}
    \centering
    \includegraphics[width=0.45\textwidth]{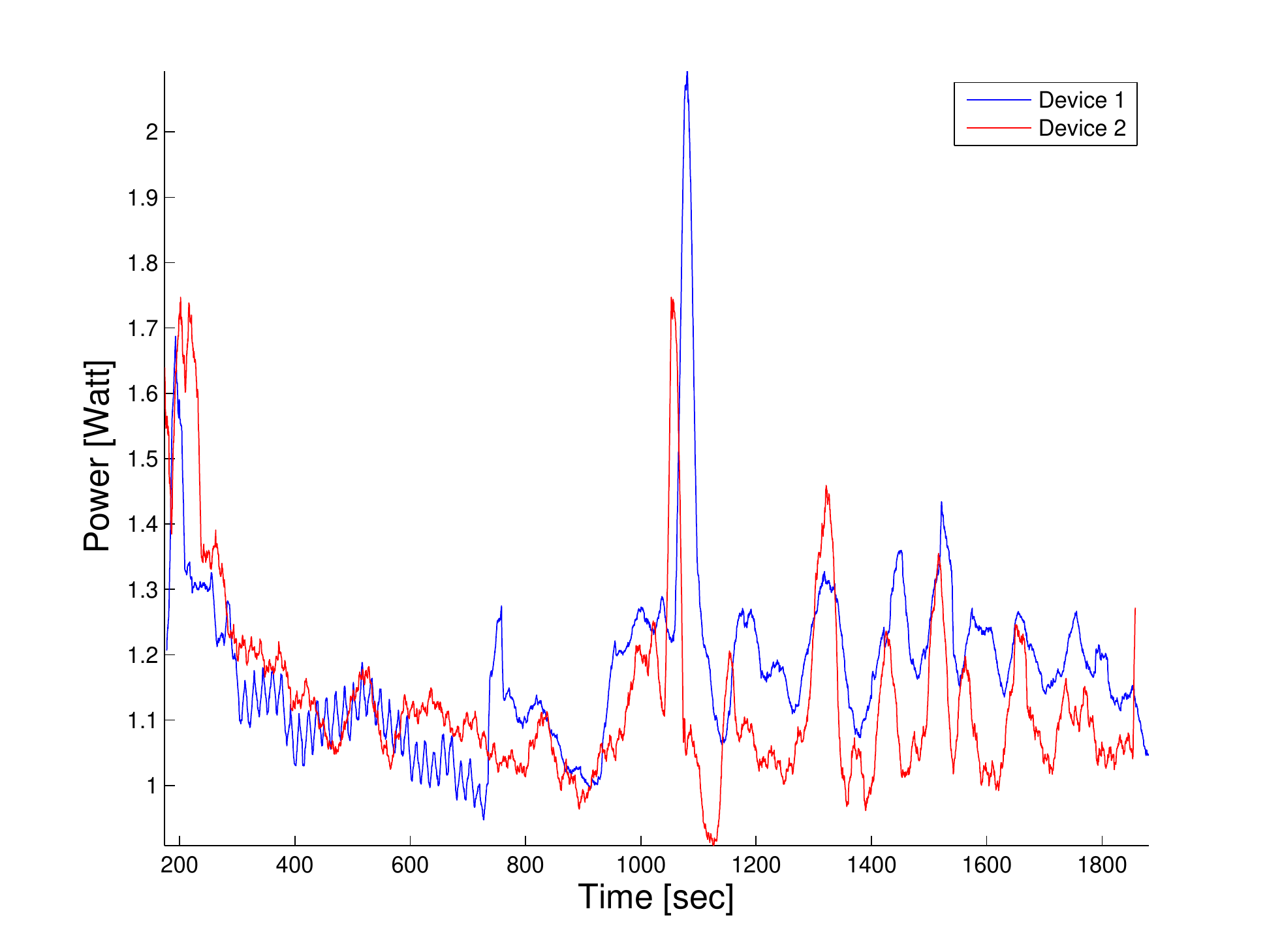}
    \caption{For two phones of the same model, power variations on the same drive are similar.}
    \label{fig:same-model-comparison}
\end{figure}

Figure~\ref{fig:same-model-comparison} shows power measurements for two Nexus~4
phones in the same vehicle, transmitting packets over their cellular link,
while driving on the same path. The power consumption variations of the Nexus~4 phones are similar,
indicating that power measurements can be mostly stable across devices.

% It should be noted that there are sections of the drive where the power
% consumption differs (e.g., 1100--1300 sec). These differences are due to the
% phones associating with different cellular base stations at that point in time.
% However, these differences do not appear to be common. We will show that power
% differences like this can be handled by using a probabilistic location
% estimation algorithm often used in robotics
% (Section~\ref{sec:particle-filter}).

\begin{figure}
    \centering
    \includegraphics[width=0.45\textwidth]{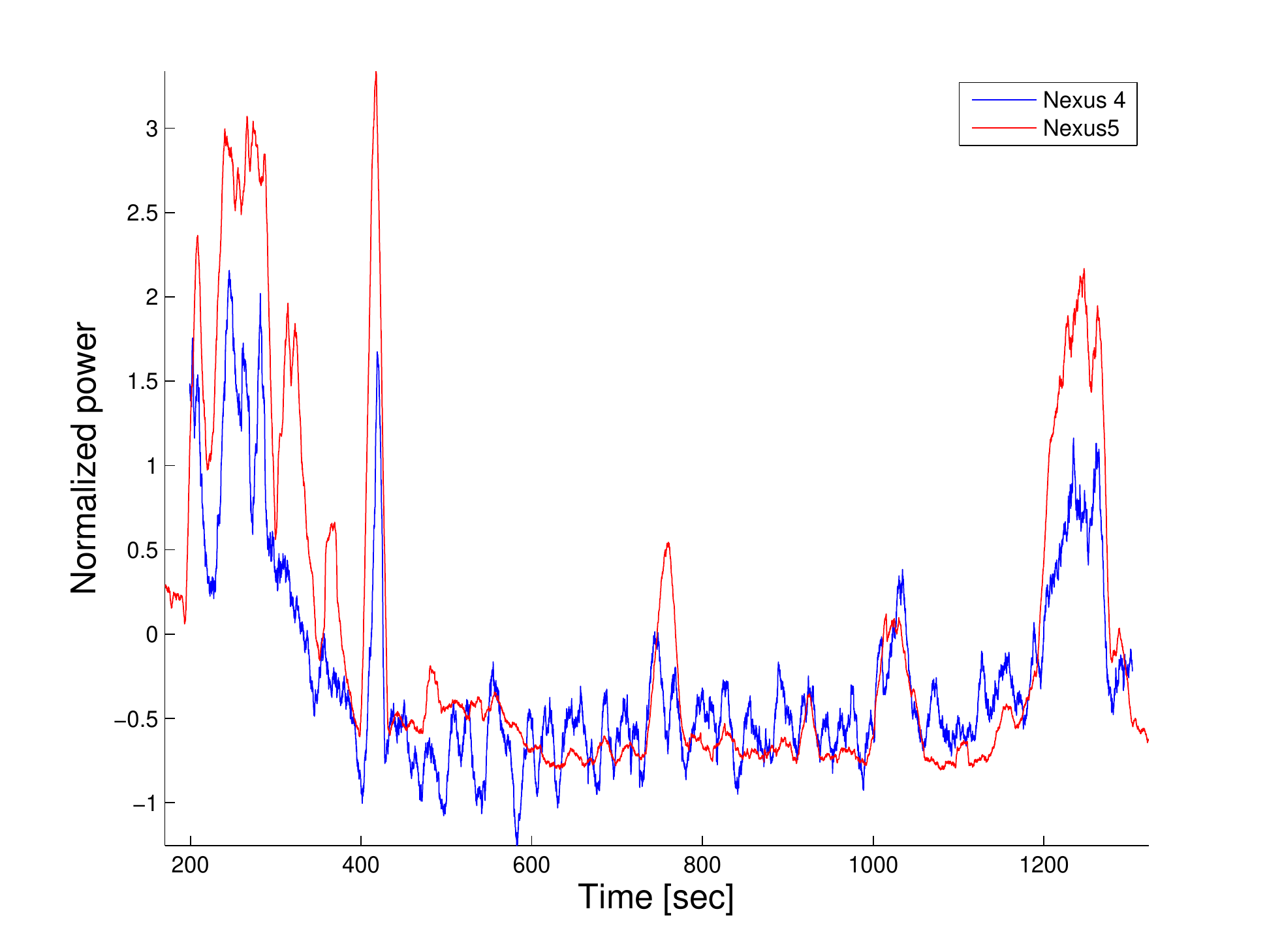}
    \caption{For two different phone models, power variations on the same drive are similar.}
    \label{fig:two-models}
\end{figure}

Finally, we demonstrate that power measurements could be stable across different
models of smartphones. This stability would allow an attacker to obtain a reference power
measurement for a drive without using the same phone as the victim's. We recorded
power measurements, while transmitting packets over cellular, using two
different smartphone models (Nexus~4 and Nexus~5) during the same ride, and we
aligned the power samples, according to absolute time.

The results presented in Figure~\ref{fig:two-models} indicate that there is
similarity between different models that could allow one model to be used as a
reference for another. This experiment serves as a proof of concept: we leave
further evaluation of such an attack scenario, where the attacker and victim use different phone models, to future work. In this paper, we assume that
the attacker can obtain reference power measurements using the same phone model as the victim.

\subsection{Hysteresis} \label{sec:hysterisis}

A phone attaches to the base station having the strongest signal. Therefore,
one might expect that the base station to which a phone is attached
and the signal strength will be the same in one location.
Nonetheless, it is shown in~\cite{Schulman2010} that signal strength can be
significantly different at a location based on how the device arrived there,
for example, the direction of arrival.  This is due to the hysteresis algorithm
used to decide when to hand-off to a new base station.  A phone hands-off from
its base station only when its received signal strength dips below the signal
strength from the next base station by more than a given
threshold~\cite{pollini1996trends}.  Thus, two phones that reside in the same
location can be attached to two different base stations.

Hysteresis has two implications for determining a victim's location from power measurements:
(1) an attacker can only use the same direction of travel as a reference power measurement,
and (2) it will complicate inferring new routes from power measurements collected from individual
road segments (Section~\ref{sec:newroutes}).

%It is tempting to reuse a profile, recorded for a certain route taken in one
%direction, as a reference profile for the opposite direction. This way the attacker can
%either decrease the number of drives and the effort it takes to map an area, or
%have more reference routes for the same amount of drives.
%
%Another implication is related to the new routes inference task.
%The hysteresis phenomena can be adversarial when trying to identify a route
%by the profiles of the invidual road segments that construct it,
%suggesting that a profile for a segment may depend on the previously taken segment.
%The relation will become more clear after section \ref{sec:newroutes},
%and as we show the effect is not strong enough to obstruct our method.

\subsection{Background summary and challenges}

The initial measurements in this section suggest that the power
consumed by the cellular radio is a side channel that leaks
information about the location of a smartphone. However, there are
four significant challenges that must be overcome to infer location
from the power meter. First, during the pre-measurement phase the
attacker may have traveled at a different speed and encountered
different stops than the target phone. Second, the attacker will have
to identify the target's power profile from among many pre-collected
power profiles along different routes. Third, once the attacker
determines the target's path, the exact location of the target on the
path may be ambiguous because of similarities in the path's power
profile. Finally, the target may travel along a path that the attacker
only partially covered during the pre-measurement phase: the attacker
may have only pre-collected measurements for a subset of segments in
the target's route. In the following sections we describe techniques
that address each of these challenges and experiment with their
accuracy.

\section{Route distinguishability}
\label{sec:route-distinguishability}

As a warm-up we show how the phone's power profile can be used to
identify what route the user is taking from among a small set of
possible routes (say, 30 routes).  Although we view it as a warm-up,
building towards our main results, route distinguishability is still
quite useful.  For example, if the attacker is familiar with the
user's routine then the attacker can pre-measure all the user's normal
routes and then repeatedly locate the user among those routes.

Route distinguishability is a classification problem: we collected
power profiles associated with known routes and want to classify new
samples based on this training set.  We treat each power profile as a
time series which needs to be compared to other time series.  A score
is assigned after each comparison, and based on these scores we select
the most likely matching route.  Because different rides along the
same route can vary in speed at different locations along the ride,
and because routes having the same label can vary slightly at certain
points (especially before getting to a highway and after exiting it),
we need to compare profile features that can vary in time and length
and allow for a certain amount of difference. We also have to
compensate for different baselines in power consumption due to
constant components that depend on the running applications and on
differences in device models.

We use a classification method based on Dynamic Time Warping (DTW)~\cite{Muller2007}, an algorithm for measuring
similarity between temporal sequences that are misaligned and vary in time or speed.
We compute the DTW distance\footnote{In fact we compute a normalized DTW distance, as we have to
compensate for difference in lengths of different routes - a longer route might yield larger DTW distance
despite being more similar to the tested sequence.} between the new power profile and all reference profiles
associated with known routes, selecting the known route that yields the minimal distance.
More formally, if the reference profiles are given by sequences $\{X\}_{i=1}^n$, and the unclassified profile
is given by sequence $Y$, we choose the route $i$ such that
$$ i = \underset{i}{\mbox{argmin}}\ \mbox{DTW}(Y, X_i)$$
which is equivalent to 1-NN classification given DTW metric.

Because the profiles might have different baselines and variability, we perform the following normalization for
each profile prior to computing the DTW distance: we calculate the mean and subtract it, and divide the result
by the standard deviation.
We also apply some preprocessing in the form of smoothing the profiles using a moving average (MA) filter in
order to reduce noise and obtain the general power consumption trend, and we downsample by a factor of
\DownsampleFactor{} to reduce computational complexity.

\section{Real-time mobile device tracking}
\label{sec:mobile-device-tracking}

In this section we consider the following task: the attacker knows
that a mobile user is traveling along a particular route and our
objective is to track the mobile device as it is moving along the
route.  We do not assume a particular starting point along the route,
meaning, in probabilistic terms, that our prior on the initial
location is uniform.  The attacker has reference power profiles
collected in advance for the target route, and constantly receives new
power measurements from an application installed on the target phone.
Its goal is to locate the device along the route, and continue
tracking it in real-time as it travels along the route.

\subsection{Tracking via Dynamic Time Warping}
\label{sec:dtw-tracking}

This approach is similar to that of route distinguishability, but we use only the measurements collected
up to this point, which comprise a sub-sequence of the entire route profile.
We use the \emph{Subsequence} DTW algorithm \cite{Muller2007}, rather than the classic DTW, to search a
sub-sequence in a larger sequence, and return a distance measure as well as the corresponding start and end
offsets.

We search for the sequence of measurements we have accumulated since the beginning of the drive in all our
reference profiles and select the profile that yields the minimal DTW distance. The location estimate
corresponds to the location associated with the end offset returned by the algorithm.

% \subsection{Motion estimation using a particle filter}
% While the previous approach alone may yield mistakes in location estimation due to a match of the measurements
% to an incorrect location, we can further improve the estimation by using a particle filter similar to \cite{Mihaylova2007}.
% However, this is beyond the scope of this paper.

\subsection{Improved tracking via a motion model}
\label{sec:improved-tracking}

While the previous approach can make mistakes in location estimation
due to a match with an incorrect location, we can further improve the
estimation by imposing rules based on a sensible motion model.  We
first need to know when we are ``locked" on the target. For this
purpose we define a similarity threshold so that if the minimal DTW
distance is above this threshold, we are in a \emph{locked} state.
Once we are locked on the target, we perform a simple sanity check at
each iteration: ``Has the target displaced by more than X?''

If the sanity check does not pass we consider the estimate unlikely to be accurate, 
and simply output the previous estimate as the new estimated location.
If the similarity is below the threshold, we switch to an \emph{unlocked} state,
and stop performing this sanity check until we are ``locked" again.
Algorithm \ref{alg:motion-model-tracking} presents this logic as pseudocode.

\begin{algorithm}
    \begin{algorithmic}
        \State $locked \gets false$ \Comment{Are we locked on the target?}
        \While{target moving}
            \State $loc[i], score \gets estimateLocation()$
            \State $d \gets getDistance(loc[i], loc[i-1])$

            \If{$locked$ and $d > MAX\_DISP$}
                \State $loc[i] \gets loc[i-1]$ \Comment{Reuse previous estimate}
            \EndIf

            \If{$score > THRESHOLD$}
                \State $locked \gets true$
            \EndIf

        \EndWhile\label{endwhile}
    \end{algorithmic}

    \caption{Improved tracking using a simple motion model}
    \label{alg:motion-model-tracking}
\end{algorithm}

\subsection{Tracking using Optimal Subsequence Bijection}
\label{sec:osb}
Optimal Subsequence Bijection (OSB) \cite{Koknar-Tezel2007} is a technique, similar to DTW, that enables aligning two sequences. In DTW, we align the query sequence with the target sequence without skipping elements in the query sequence, thereby assuming that the query sequence contains no noise. OSB, on the other hand, copes with noise in both sequences by allowing to skip elements. A fixed jump-cost is incurred with every skip in either the query or the target sequence. This extra degree of freedom has potential for aligning noisy subsequences more efficiently in our case.
In the evaluation section we present results obtained by using OSB and compare them to those obtained using DTW.

\section{Inference of new routes} \label{sec:newroutes}
In Section~\ref{sec:route-distinguishability} we addressed the problem of identifying the route traversed
by the phone, assuming the potential routes are known in advance.
This assumption allowed us to train our algorithm specifically for the potential routes.
As previously mentioned, there are indeed many real-world scenarios where it is applicable.
Nevertheless, in this section we set out to tackle a broader tracking problem, where the future
potential routes are not explicitly known. Here we specifically aim to identify the final location of the phone after it traversed an unknown route.
We assume that the area in which the mobile device owner moves is known,
however the number of all possible routes in that area may be too large to practically pre-record each one.
Such an area can be, for instance, a university campus, a neighborhood, a small town or a highway network.

We address this problem by pre-recording the power profiles of all the road segments within the given area.
Each possible route a mobile device may take is a concatenation of some subset of these road segments.
Given a power profile of the tracked device, we will reconstruct the unknown route using the reference power
profiles corresponding to the road segments. The reconstructed route will enable us to estimate the phone's final location.
Note that, due to the hysteresis of hand-offs between cellular base stations, a
power consumption is not only dependent on the traveled road segment, but also on the previous road
segment the device came from.

In Appendix~\ref{sec:model} we formalize this problem as a hidden Markov model (HMM) \cite{Rabiner1989}.
In the following we describe a method to solve the problem using a particle filter.
The performance of the algorithm will be examined in the next section.

\subsection{Particle Filter} \label{sec:particle-filter}
A particle filter~\cite{arulampalam2002tutorial} is a method that estimates the state of a HMM at each step based
on observations up to that step.
The estimation is done using a Monte Carlo approximation where a set of samples (particles) is generated at each
step that approximate the probability distribution of the states at the corresponding steps.
A comprehensive introduction to particle filters and their relation to general state-space models is provided in
\cite{ristic2004beyond}.

We implement the particle filter as follows. We denote $O^r=\left\{ o^r_{xyz} \right\}$, where $o^r_{xyz}$ is a power profile prerecorded over segment $(y,z)$ while the segment $(x,y)$ had been traversed just before it. We use a discrete time resolution $\tau=3$ seconds. We denote $\Delta^{yz}_{\min}$ and $\Delta^{yz}_{\max}$ to be the minimum and maximum time duration to traverse road segment $(y,z)$, respectively. We assume these bounds can be derived from prerecordings of the segments. At each iteration $i$ we have a sample set of $N$ routes $P_i=\left\{(Q,T)\right\}$. The initial set of routes $P_0$ are chosen according to $\Pi$.
At each step, we execute the following algorithm:

\begin{algorithm}
    \begin{algorithmic}
		\ForAll{route $p$ in $P$}
			\State $t_{\textrm{end}} \gets$ end time of $p$
			\State $(x,y) \gets$ last segment of $p$
			\State $z \gets$ next intersection to traverse (distributed by $A$)
			\vspace{-15px}
			\State \[\hspace{-6px}W_p \gets \min_{\begin{subarray}{c}t \in [\Delta^{yz}_{\min}, \Delta^{yz}_{\max}]\\ o^r_{xyz} \in O^r_{xyz}\end{subarray}} \left\{\textrm{DTW}(O_{[t_{\textrm{end}},t_{\textrm{end}}+t]}, o^r_{xyz})\right\}
				\]
				\vspace{-9px}
				\State $p \gets p || (y,z)$
				\State Update the end time of $p$
		\EndFor
		\State Resample $P$ according to the weights $W_p$
    \end{algorithmic}

    \caption{Particle filter for new routes estimation}
    \label{alg:new-route-particle-filter}
\end{algorithm}

At each iteration, we append a new segment, chosen according to the prior $A$, to each possible route
(represented by a particle).
Then, the traversal time of the new segment is chosen so that it will have a minimal DTW
distance to the respective time interval of the tracked power profile. We take this minimal distance as the weight
of the new route. After normalizing the weights of all routes, a resampling phase takes place.
$N$ routes are chosen from the existing set of routes according to the particle weights distribution\footnote{Note that the resampling of the new routes can have repetitions.
Namely, the same route can be chosen more than one time}.
The new resampled set of routes is the input to the next iteration of the particle filter.
The total number of iterations should not exceed an upper bound on the number of segments that the tracked
device can traverse. Note however that a route may exhaust the examined power profile before the last iteration (namely, the end time of that route reached $t_{\max}$). In such a case we do not update the route in all subsequent iterations (this case is not described in Algorithm~\ref{alg:new-route-particle-filter} to facilitate fluency of exposition).

Before calculating the DTW distance of a pair of power profiles the profiles are preprocessed to remove as much noise as possible.
We first normalize the power profile by subtracting its mean and dividing by the standard deviation
of all values included in that profile. Then, we zero out all power values below a threshold
percentile. This last step allows us to focus only on the peaks in power consumption where the radio's power
consumption is dominant while ignoring the lower power values for which the radio's power has a lesser effect.
The percentile threshold we use in this paper is 90\%.

Upon its completion, the particle filter outputs a set of $N$ routes of various lengths.
To select the best estimate route the simple approach is to choose the route that appears the most number of times in the output set as it has the highest probability to occur. Nonetheless, since a route is composed of multiple segments chosen at separate steps, at each step the weight of a route is determined solely based on the last segment added to the route. Therefore, the output route set is biased in favor of routes ending with segments that were given higher weights, while the weights of the initial segments have a diminishing effect on the route distribution with every new iteration.
To counter this bias, we choose another estimate route using a procedure we call \emph{iterative majority vote}, described is Appendix~\ref{sec:bestroute}.

\section{Experiments} \label{sec:Experiments}

\subsection{Data collection} \label{sec:data-collection}
Our experiments required collecting real power consumption data from smartphone devices along
different routes. We developed the PowerSpy android application\footnote{Source code can be obtained from \newline
  \textcolor{blue}{\url{https://bitbucket.org/ymcrcat/powerspy}}.}
that collects various
measurements including signal strength, voltage, current, GPS coordinates, temperature,
state of discharge (battery level) and cell identifier.
The recordings were performed using Nexus 4, Nexus 5 and HTC mobile devices.

\subsection{Assumptions and limitations}

Exploring the limits of our attack, i.e.\ establishing the minimal necessary conditions for it to work,
is beyond our resources. For this reason, we state the assumptions on which we rely in our methods.

We assume there is enough variability in power consumption along a route to exhibit unique features.
Lack of variability may be due to high density of cellular antennas that flatten the signal strength profile.
We also assume that enough communication is occurring for the signal strength to have an effect on
power consumption.
This is a reasonable assumption, since background synchronization of data happens frequently in
smartphone devices. Moreover, the driver might be using navigation software or streaming music.
However, at this stage, it is difficult to determine how inconsistent phone usage across different rides will
affect our attacks.

Identifying which route the user took involves understanding which
power measurements collected from her mobile device occurred during
driving activity.  Here we simply assume that we can identify driving
activity.  Other works (e.g.,~\cite{Mohan2008}) address this question
by using data from other sensors that require no permissions, such as
gyroscopes and accelerometers.

\begin{figure}
    \centering
    \includegraphics[width=0.4\textwidth]{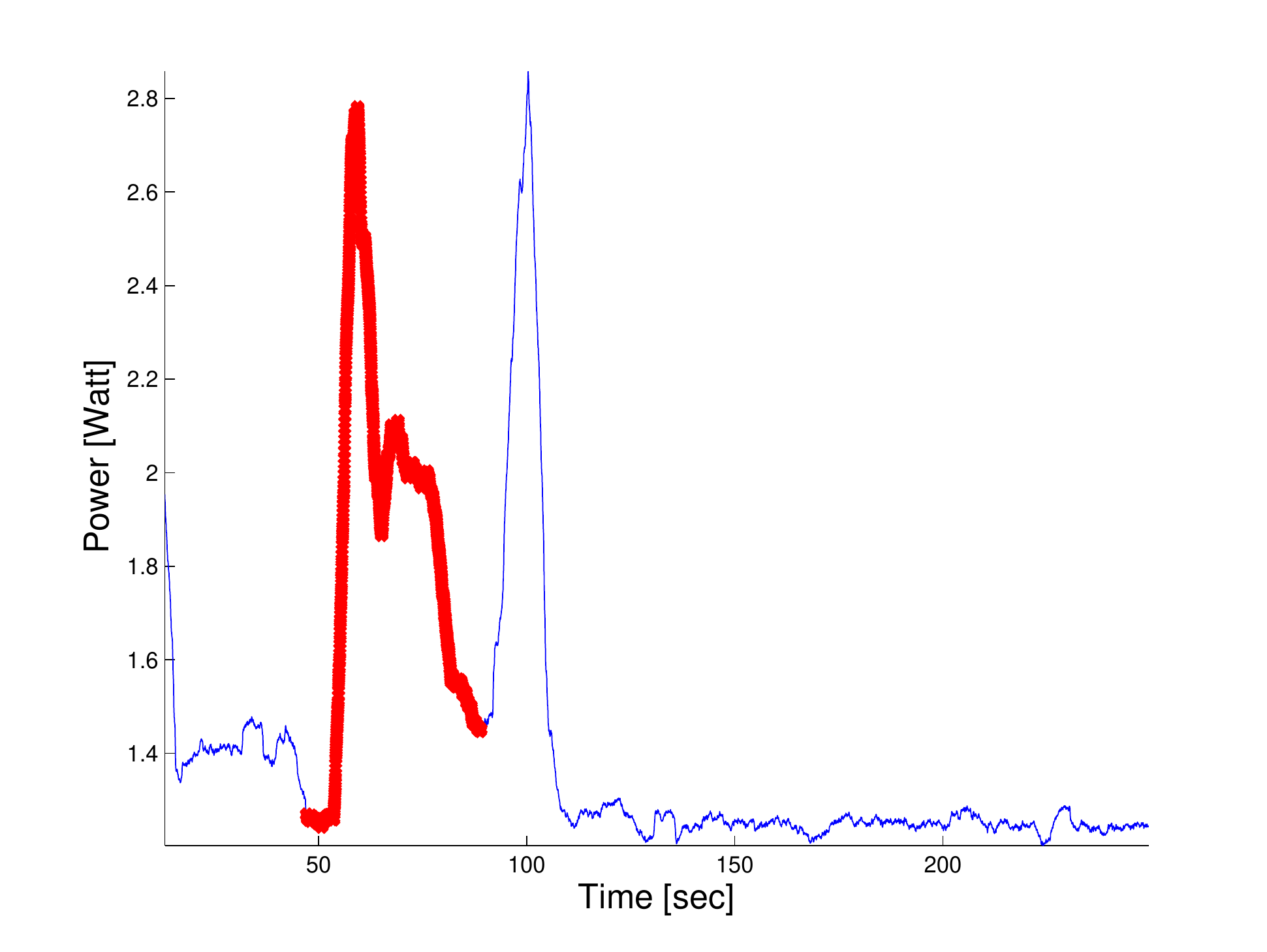}
    \caption{Power profile with a phone call occurring between 50-90 seconds.
        Profile region during phone call is marked in red.}
    \label{fig:phone-call}
\end{figure}

Some events that occur while driving, such as an incoming phone call,
can have a significant effect on power consumption.
Figure~\ref{fig:phone-call} shows the power profile of a device at
rest when a phone call takes place (the part marked in red).  The peak
immediately after the phone call is caused by using the phone to
terminate the phone call and turn off the display.  We can see that
this event appears prominently in the power profile and can cope with
such transient effects by identifying and truncating peaks that stand
out in the profile.  In addition, smoothing the profile by a moving
average should mitigate these transient effects.

\subsection{Route distinguishability}
To evaluate the algorithm for distinguishing routes (\cref{sec:route-distinguishability})
we recorded reference profiles for multiple different routes.
The profiles include measurements from both Nexus 4 and Nexus 5 models.
In total we had a dataset of 294 profiles, representing 36 unique routes.
Driving in different directions along the same roads (from point A to B vs. from point B to A)
is considered two different routes.
We perform cross validation using multiple iterations (100 iterations),
each time using a random portion of the profiles as a training set, and requiring
equal number of samples for each possible class.
The sizes of the training and test sets depend on how many reference routes per profile we require
each time. Naturally, the more reference profiles we have, the higher the identification rate.

One evaluation round included 29 unique routes, with only 1 reference profile per route in the training
set, and 211 test routes. It resulted in correct identification rate of 40\%.
That is compared to the random guess probability of only 3\%.
Another round included 25 unique routes, with 2 reference profiles per route in the training set
and 182 routes in the test set, and resulted in correct identification rate of 53\%
(compared to the random guess probability of only 4\%).
Having 5 reference profiles per route (for 17 unique routes) raises the identification rate to 71\%,
compared to the random guess probability of 5.8\%. And finally, for 8 reference profiles per route we get
85\% correct identification.
The results are summarized in \cref{tab:route-distinguishability-eval}.

We can see that an attacker can have a significant advantage in guessing the route taken
by a user.

\begin{table*}
    \centering
    \begin{tabular}{cccccc}
     {\bf \# Unique Routes} &  {\bf \# Ref. Profiles/Route} &  {\bf \# Test Routes} &
     {\bf Correct Identification \%} &  {\bf Random Guess \%} \\
        \hline
          8 &           10 &          55 &       85 &           13 \\
         17 &            5 &         119 &       71 &            6 \\
         17 &            4 &         136 &       68 &            6 \\
         21 &            3 &         157 &       61 &            5 \\
         25 &            2 &         182 &       53 &            4 \\
         29 &            1 &         211 &       40 &            3 \\
    \end{tabular}
    \caption{Route distinguishability evaluation results. First column indicates the number of
    unique routes in the training set. Second column indicates the number of training samples per route
    at the attacker's disposal. Number of test routes indicates the number of power profiles
    the attacker is trying to classify.
    Correct identification percentage indicates the percentage of correctly identified routes as
    a fraction of the third column (test set size),
    which could be then compared to the expected success of random guessing in
    the last column.}
    \label{tab:route-distinguishability-eval}
\end{table*}

% We also evaluated the algorithm on a dataset of 18 profiles for 2 different routes of about 20 kilometers,
% collected in a completely different area\footnote{Different country and a different cellular provider.} with higher
% cell density and obtained somewhat lower correct classification rate of of 78\%, which is nevertheless
% significantly better than a random guess. We attribute the decrease in correct classification to
% higher cell density, resulting in more monotonous power profiles.

\subsection{Real-time mobile device tracking}
We evaluate the algorithm for real-time mobile device tracking (\cref{sec:mobile-device-tracking})
using a set of 10 training profiles and an additional test profile.
The evaluation simulates the conditions of real-time
tracking by serially feeding samples to the algorithm as if they are received
from an application installed on the device.
We calculate the estimation error, i.e.\ the distance between the estimated coordinates
and the true location of the mobile device at each step of the simulation.
We are interested in the \emph{convergence time}, i.e.\ the number of samples it takes
until the location estimate is close enough to the true location, as well as in the distribution
of the estimation errors given by a histogram of the absolute values of the distances.

\begin{figure*}
  \centering
  \begin{subfigure}{0.4\textwidth}
    \includegraphics[width=\textwidth]{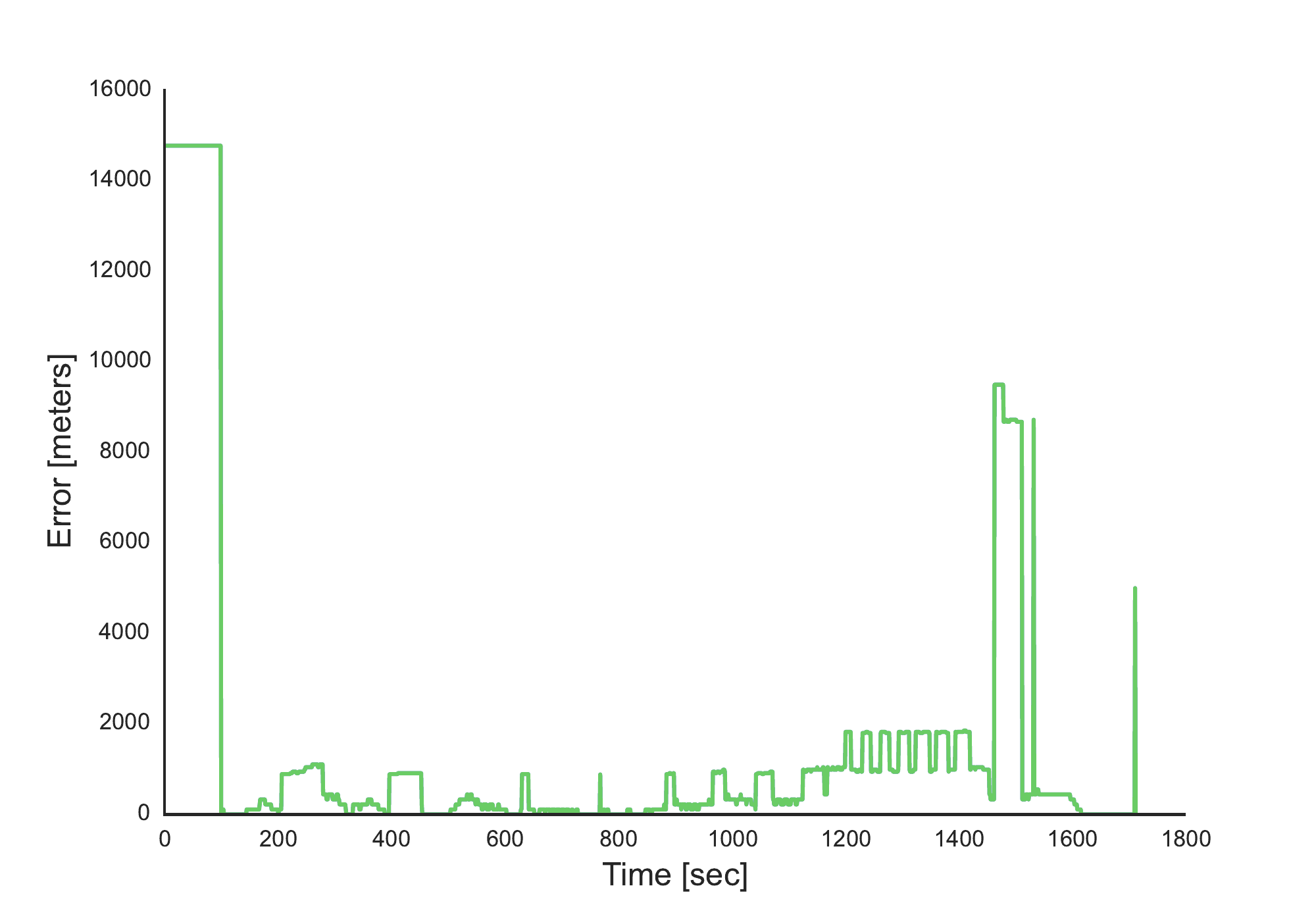}
    \caption{Convergence to true location.}
    \label{fig:est-error}
  \end{subfigure}
  \hspace{2cm}
  \begin{subfigure}{0.4\textwidth}
      \includegraphics[width=\textwidth]{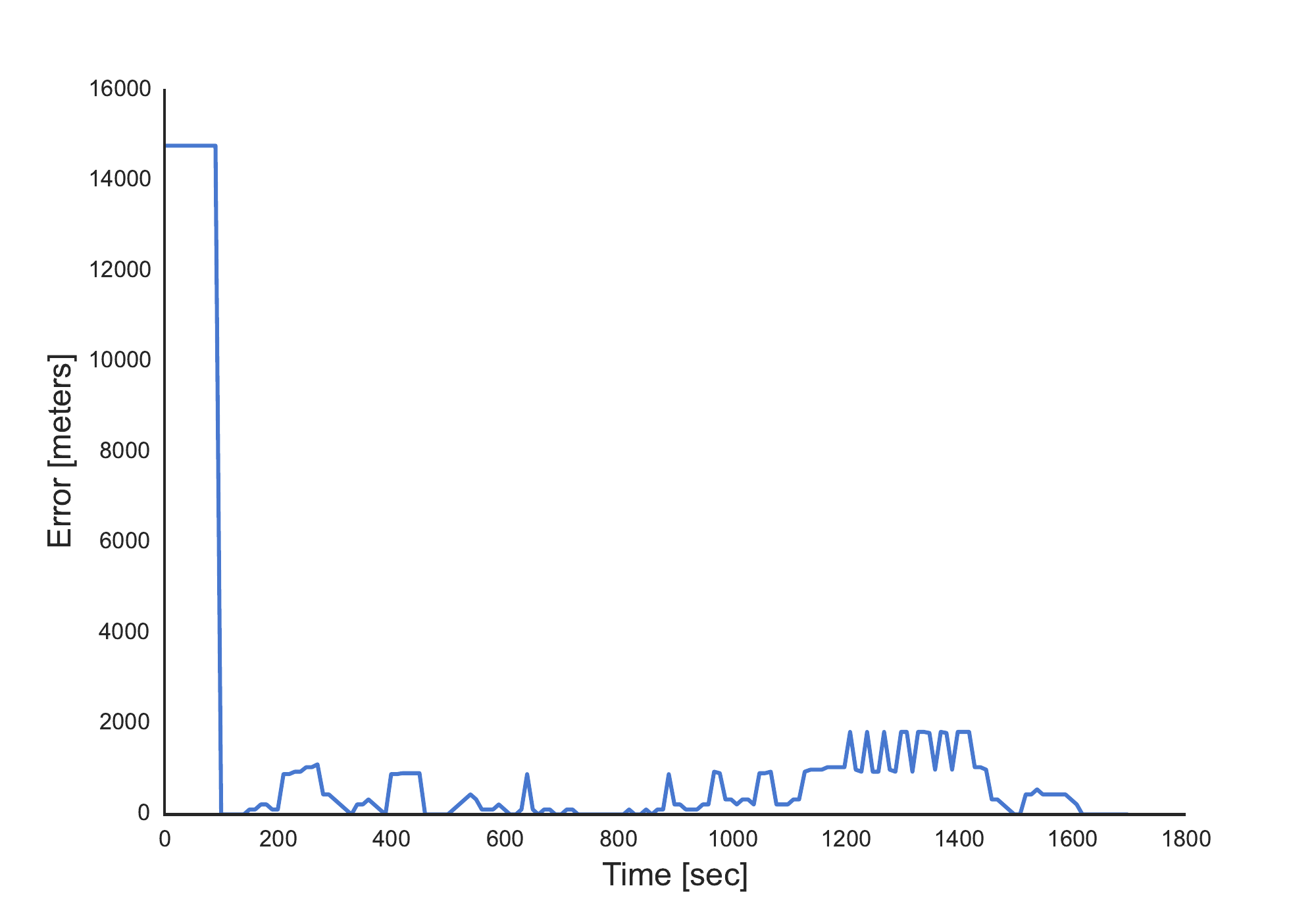}
      \caption{Location estimation error for improved tracking algorithm.}
      \label{fig:tracking-error-improved}
  \end{subfigure}

  \caption{Location estimation error for online tracking.}
  \label{fig:tracking-error}
\end{figure*}

Figure \ref{fig:tracking-error} illustrates the performance of our tracking algorithm for
one of the routes, which was about 19 kilometers long.
At the beginning, when there are very few power samples, the location estimation is extremely inaccurate,
but after two minutes we lock on the true location.
We obtained a precise estimate from 2 minutes up until 20 minutes on the route,
where our estimate slightly diverges, due to increased velocity on a freeway segment.
Around 26 minutes (in figure \ref{fig:est-error}) we have a large estimation error,
but as we mentioned earlier, these kind of errors are easy to prevent by imposing a simple motion model
(\cref{sec:improved-tracking}).
Most of the errors are small compared to the length of the route: 80\% of the estimation errors are less than 1 km.

\begin{figure*}
    \centering

    \begin{subfigure}{0.4\textwidth}
      \includegraphics[width=\textwidth]{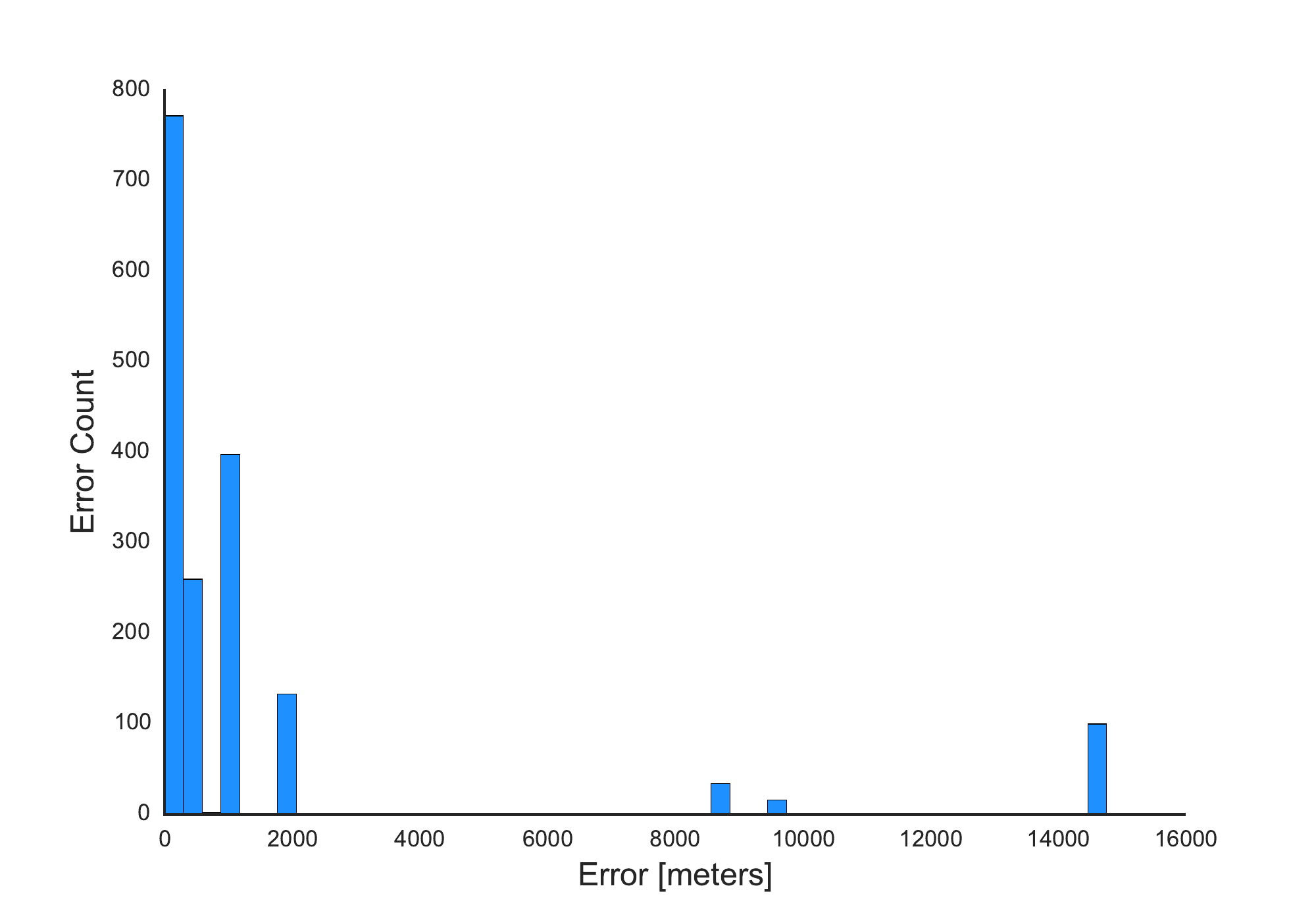}
      \caption{Errors histogram. Almost 90\% of the errors are less than 1 km.}
    \end{subfigure}
    \hspace{2cm}
    \begin{subfigure}{0.4\textwidth}
      \includegraphics[width=\textwidth]{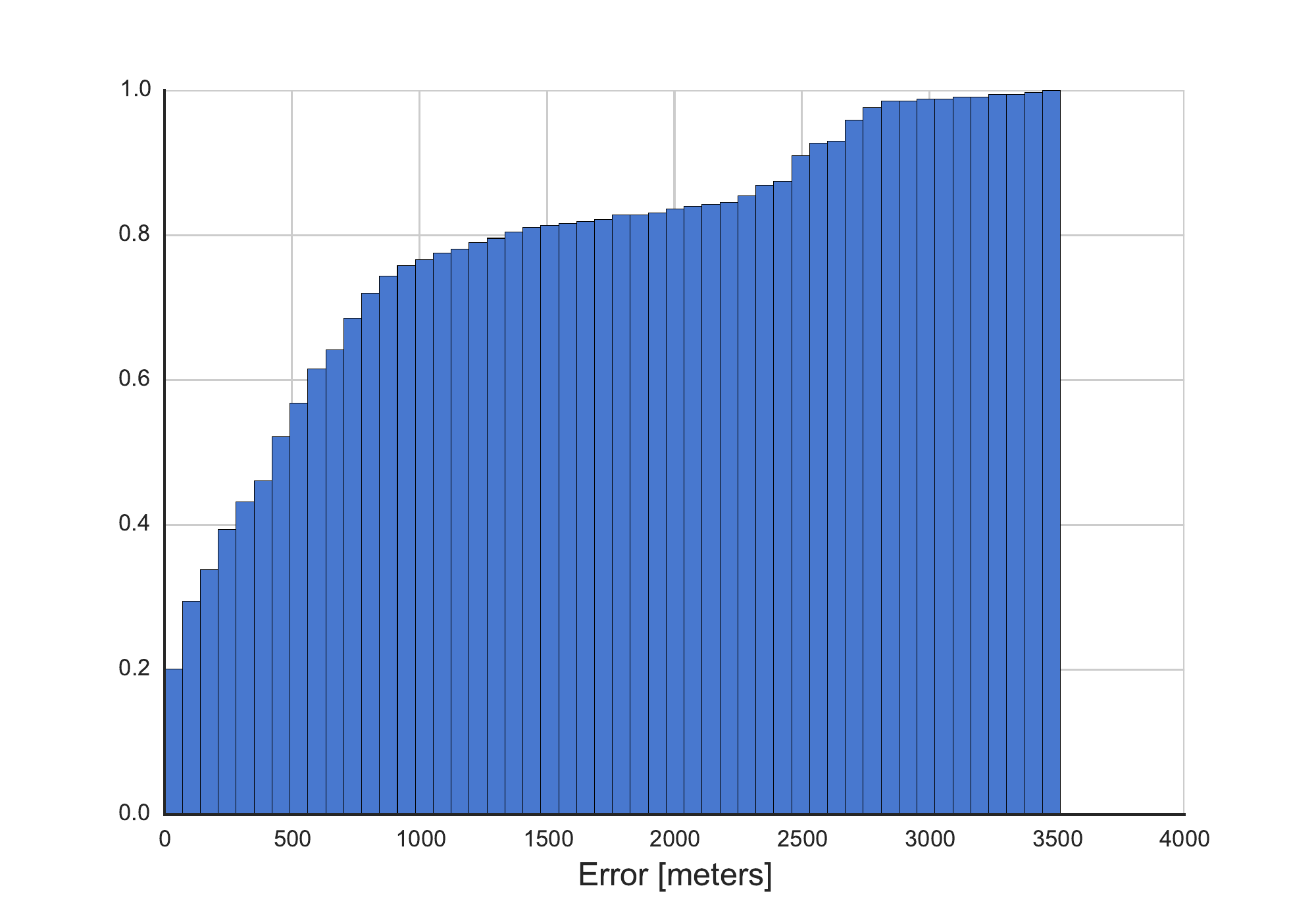}
      \caption{Error cumulative distribution.}
    \end{subfigure}

    \caption{Estimation errors distribution for motion-model tracking.}
    \label{fig:error-distribution}
\end{figure*}

We also tested the improved tracking algorithm explained in \cref{sec:improved-tracking}.
Figure \ref{fig:tracking-error-improved} presents the estimation error over time,
and we can see that the big errors towards the end of the route that appeared in
\ref{fig:est-error} are not present in \cref{fig:tracking-error-improved}. Moreover, now almost 90\% of the estimation
errors are below 1 km (\cref{fig:error-distribution}).

We provide animations visualizing our results for real-time tracking at the following links.
The animations, generated using our estimations of the target's location, depict
a moving target along the route and our estimation of its location.
The first one corresponds to the method described in \ref{sec:dtw-tracking},
and the second to the one described in \ref{sec:improved-tracking} that uses the
motion model based correction: \newline
\textcolor{blue}{\url{crypto.stanford.edu/powerspy/tracking1.mov}} \newline
\textcolor{blue}{\url{crypto.stanford.edu/powerspy/tracking2.mov}}

\subsubsection{OSB vs. DTW}
\begin{figure}
   \centering
   \begin{subfigure}{0.4\textwidth}
        \includegraphics[width=\textwidth]{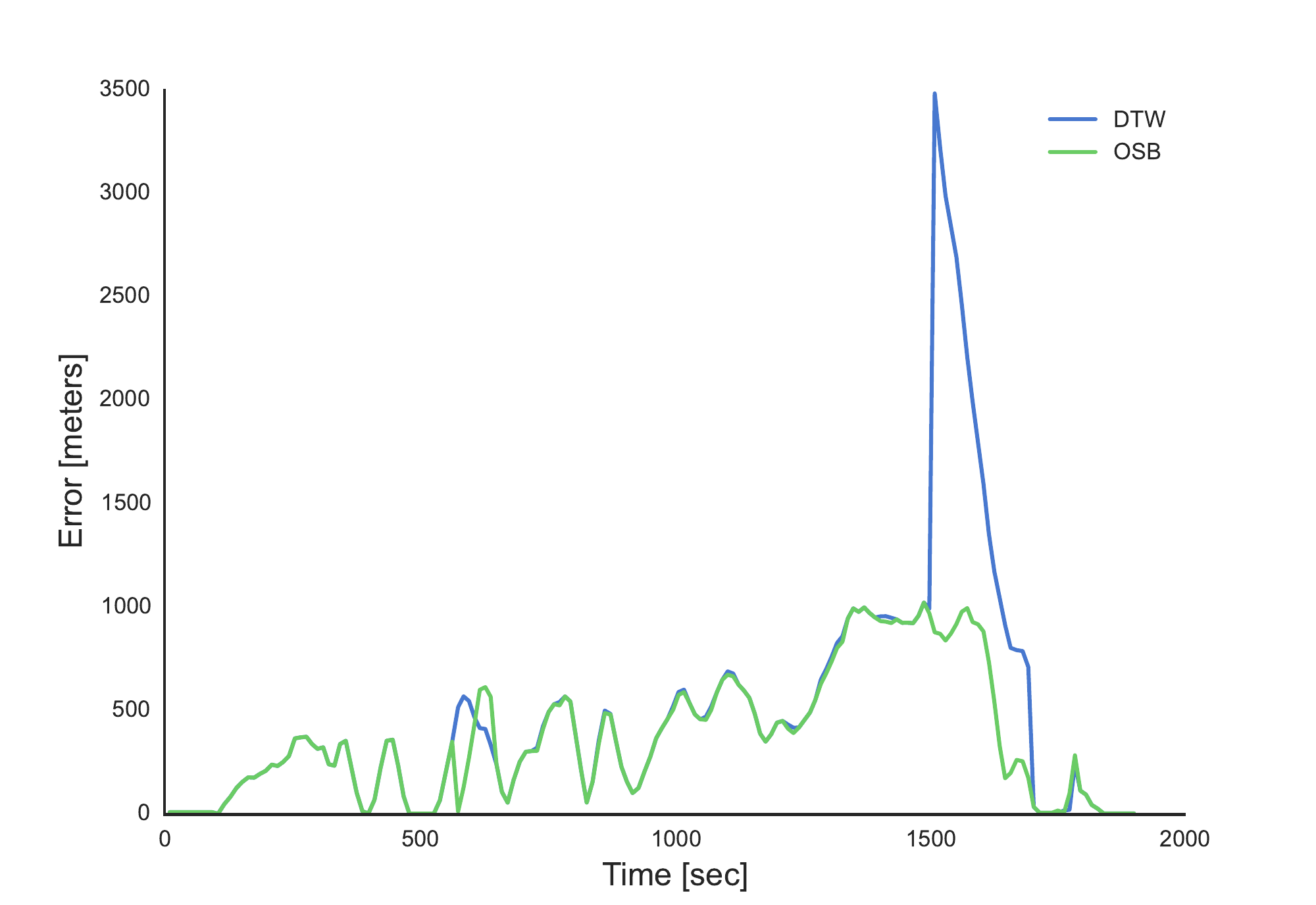}
   \end{subfigure}
   \begin{subfigure}{0.4\textwidth}
        \includegraphics[width=\textwidth]{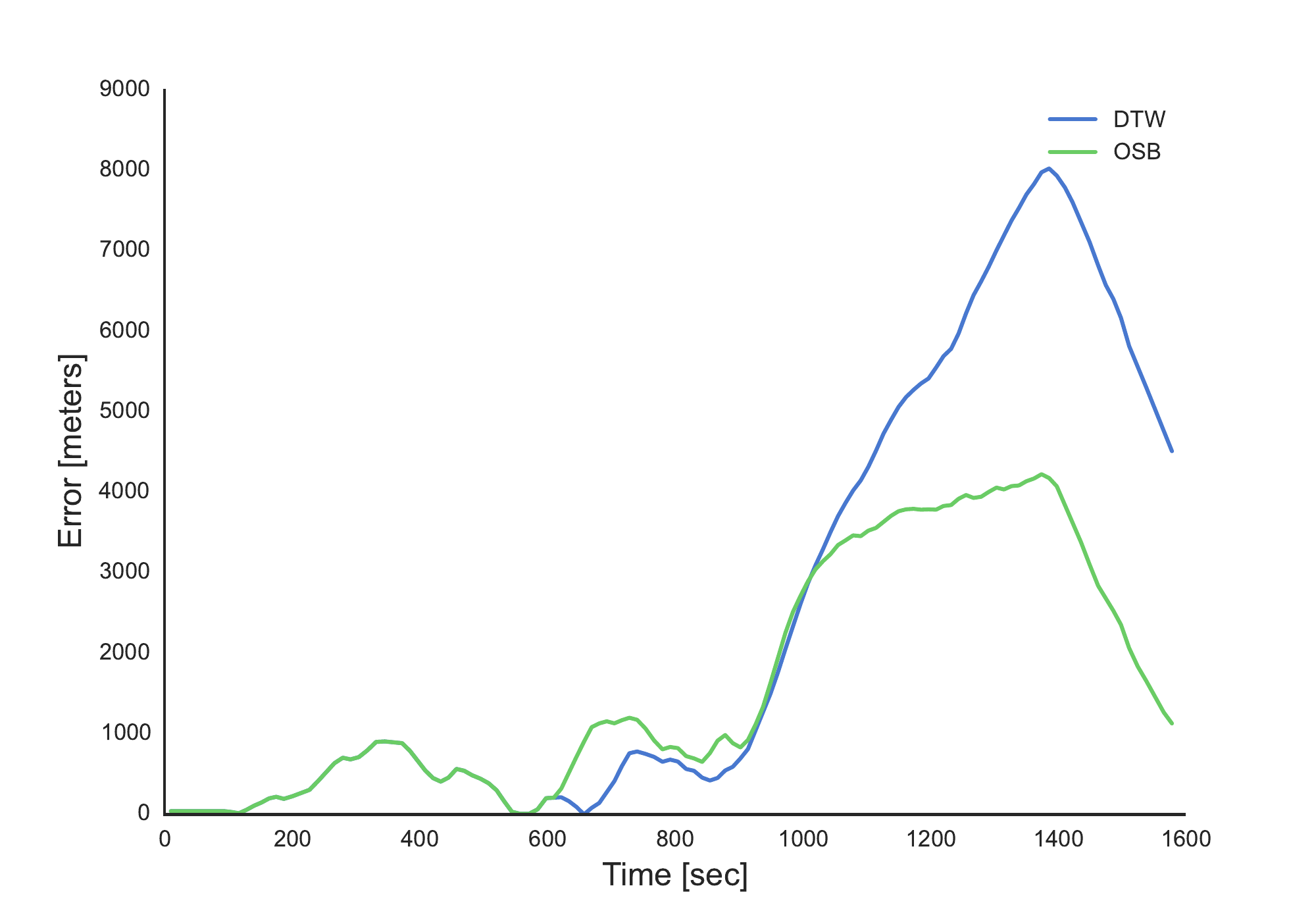}
   \end{subfigure}
   \caption{Comparison of DTW and OSB for real-time tracking.}
   \label{fig:dtw_vs_osb}
\end{figure}

We compare the performance of Dynamic Time Warping to that of Optimal Subsequence Bijection (\cref{sec:osb}).
Figure \ref{fig:dtw_vs_osb} present such a comparison for the same route, using two different recordings.
The tracking was performed without compensating for errors using a motion model, to evaluate the performance
of the subsequence matching algorithms as they are.
We can see that, in both cases, Optimal Subsequence Bijection outperforms the standard Subsequence-DTW most of the time.
Therefore, we suggest that further experimentation with OSB could potentially be beneficial for this task.

\subsection{Inference of new routes}
\subsubsection{Setup}

\begin{figure}
    \centering
    \includegraphics[width=0.35\textwidth]{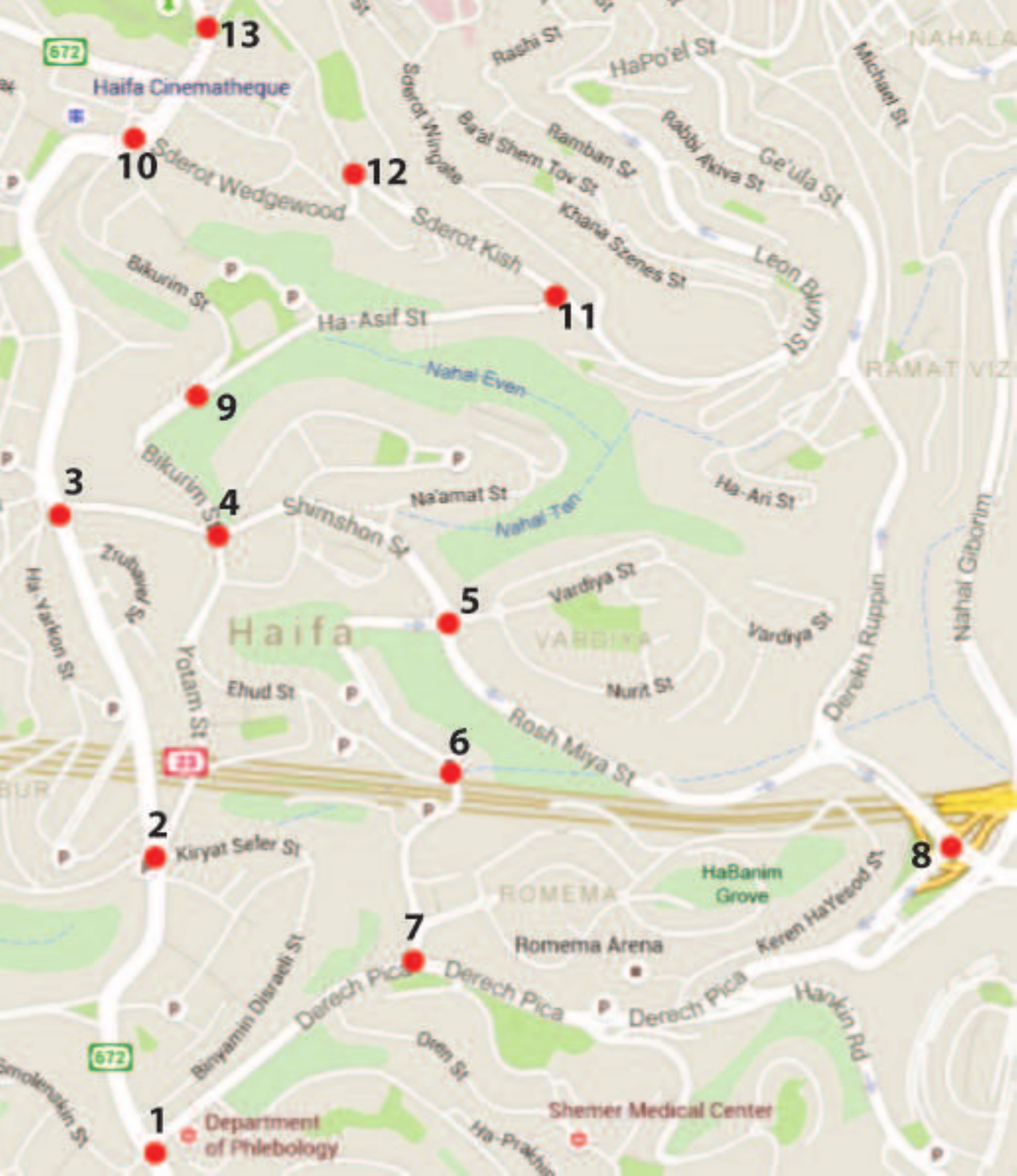}
    \caption{Map of area and intersections for route inference.}
    \label{fig:area}
\end{figure}

For the evaluation of the particle filter presented in Section~\ref{sec:newroutes} we considered an area
 depicted in Figure~\ref{fig:area}. The area has 13 intersections
having 35 road segments\footnote{Three of the segments are one way streets.}. The
average length of a road segment is about 400 meters. The average travel time over the segments is around 70
seconds. The area is located in the center of Haifa, a city located in northern Israel, having a population density comparable to Philadelphia or Miami. 
Traffic congestion in this area varies across segments and time of day.
For each power recording, the track traversed at least one congested segment.
Most of the 13 intersections have traffic lights, and about a quarter of the road segments pass through them.

We had three pre-recording sessions which in total covered all segments. Each road segment was entered from every possible direction to account for the hysteresis effects.
The pre-recording sessions were done using the same Nexus 4 phone.

We set the following  parameters of the HMM (as they are defined in Appendix~\ref{sec:model}):
\begin{enumerate}
    \item $A$ -- This set defines the transition probabilities between the road segments.
    We set these probabilities to be uniformly distributed over all possible transitions.
    Namely, $a_{xyz} = \left\{ 1/|I_y| ~ | I_y = \left\{ w | (y,w)\in R, w \neq x  \right\} \right\}$.
    \item $B$ -- This set defines the distribution of power profile observations over each state.
    These probabilities depend on the road segments and their location relative to the nearby based stations.
    We do not need an explicit formulation of these probabilities to employ the particle filter.
    The likelihood of a a power profile to be associated with a road segment is estimated by the DTW distance of
    the power profile to prerecorded power profiles of that segment.
    \item $\Pi$ -- This set defines the initial state distribution.
    We assume that the starting intersection of the tracked device is known.
    This applies to scenarios where the tracking begins from well-known locations, such as the user's home,
    office, or another location the attacker knows in advance.
\end{enumerate}

For testing, we used 4 phones: two Nexus 4 (different from the one used for the pre-recordings), a \mbox{Nexus 5} and an HTC Desire.
Each phone was used to record the power profile of a different route.
The four routes combined cover almost all of the road segments in the area.
Table~\ref{tab:TestRoutes} details the routes by their corresponding sequences of intersection identifiers. 
These route recordings were done on different days, different time of day and varying weather conditions.

As noted, we can only measure the aggregate power consumption which can be significantly affected by applications that run continuously. 
To have a better sense of the effects of these applications the phones were run with different number of background applications. \mbox{Nexus 4 \#1}, 
\mbox{Nexus 5} and HTC Desire have a relatively modest number of applications which included (beyond the default Android apps): Email (corporate account), Gmail, and Google Calender. Nexus 4 \#2 has a much higher number of application which included on top of the applications of phone \#1: Facebook, Twitter, Skype, Waze, and WhatsApp. All those applications periodically send and receive traffic.

\begin{table}
	\centering
	\small
	\begin{tabular}{|c|c|}
		\hline
		Phone & Track \\
		\hline
		Nexus 4 \#1 & 8-5-6-7-1-2-3-4-5-6-4-3-2-1-7-8\\
		\hline
		Nexus 4 \#2 & 7-1-2-3-4-5-8-7-6-5-4-2-1-7-8\\
		\hline
		Nexus 5 & 3-2-4-9-10-12-11-9-4-5-6-4-3-2-1-7-6-5-8-7\\
		\hline
		HTC Desire & 10-12-11-9-4-2-1-7-6-5-8\\
    \hline
	\end{tabular}
	\normalsize
	\caption{Test Routes}
	\label{tab:TestRoutes}
\end{table}

For each of the four tracks we derived all possible sub-tracks having 3 to 7 road segments. We estimated each such sub-track.
In total we estimated around 200 sub-tracks. For each sub-track we
employed Algorithms~\ref{alg:new-route-particle-filter} and \ref{alg:iterative-majority-vote} to get two best
estimates for the sub-track.

\Cref{tab:DestinationLocalization,tab:LevenshteinDistance,tab:ExactFullRouteFit} summarize the results of route estimation for each of
the four phones. For each route we have two alternatives for picking an estimate (1) the most frequent route in
the particle set as output by Algorithm~\ref{alg:new-route-particle-filter}; (2) the route output by Algorithm~\ref{alg:iterative-majority-vote}.
For each alternative we note the road segment in which the phone is estimated to be after the completion of its track and compare it with the final road segment of the true route. This allows us to measure the accuracy of the algorithm for estimating the location of the user's destination (the end of the track). This is the most important metric for many attack scenarios where the attacker wishes to learn the destination of the victim.

In some cases it may also be beneficial for the attacker to know the actual route through which the victim traversed on his way to the destination.
For this purpose, we also calculate for each alternative estimate the Levenshtein distance between it and the true route. The Levenshtein distance is a standard metric for measuring the difference between two sequences~\cite{levenshtein1966binary}. It equals the minimum number of updates required in order to change one sequence to the next.
In this context, we treat a route as a sequence of intersections.
The distance is normalized by the length of the longer route of the two. This allows us to measure the accuracy of the algorithm for estimating the full track the user traversed.
For each estimate we also note whether it is an exact fit with the true route (i.e., zero distance).
The percentage of successful localization of destination, average Levenshtein distance and percentage of exact full route fits are calculated for each type of estimated route.
We also calculate these metrics for both estimates combined while taking into account for each track the best of the two estimates.
To benchmark the results we note in each table the performance of a random estimation algorithm which simply outputs a random, albeit feasible, route.

\begin{table}
	\centering
	\small
		\begin{tabular}{c|c|c|c|c|}
			\cline{2-5}
			& random & frequent & Alg.~\ref{alg:iterative-majority-vote} & combined \\
			\hline
			\multicolumn{1}{|c|}{Nexus 4 \#1} & 33\% & 65\% & 48\% & 80\% \\
			\hline
			\multicolumn{1}{|c|}{Nexus 4 \#2} & 31\% & 48\% & 56\% & 72\% \\
			\hline
			\multicolumn{1}{|c|}{Nexus 5} & 20\% & 33\% & 32\% & 55\% \\
			\hline
			\multicolumn{1}{|c|}{HTC Desire} & 22\% & 40\% & 41\% & 65\% \\
			\hline
		\end{tabular}
		\normalsize
	\caption{Destination localization}
	\label{tab:DestinationLocalization}
\end{table}

The results in Table~\ref{tab:DestinationLocalization} show the accuracy of destination identification. It is evident that the performance of the most frequent
route output by the particle filter is comparable to the performance of the best estimate output by
Algorithm~\ref{alg:iterative-majority-vote}.
However, their combined performance is significantly better than either estimates alone and predict more accurately the final destination of the phone.
This result suggests that Algorithm~\ref{alg:iterative-majority-vote} extracts significant amount of information from
the routes output by the particle filter beyond the information gleaned from the most frequent route.

Table~\ref{tab:DestinationLocalization}  indicates that for Nexus 4 \#1 the combined route estimates were able to identify the final road segment for
80\% of all scenarios. For Nexus 4 \#2 which was running many applications the final destination  estimates are somewhat less accurate (72\%). 
This is attributed to the more noisy measurements of the aggregate power consumption. 
The accuracy for the two models -- Nexus 5 and HTC Desire -- is lower than the accuracy achieved for Nexus 4. Remember that all our pre-recordings were done using a Nexus 4. These results may indicate that the power consumption profile of the cellular radio is dependent on the phone's model. Nonetheless, for both phones we achieve significantly higher accuracy of destination localization (55\% and 65\%) as compared to the random case (about 20\%).

\Cref{tab:LevenshteinDistance,tab:ExactFullRouteFit} present measures -- Levenshtein distance and exact full route fit -- of the accuracy of estimates for the full route the phone took to its destination. Here, again, the algorithm presented for Nexus 4 \#1 superior performance. It was able to exactly estimate 45\% of the full route to the destination. On the other hand, for the more busy Nexus 4 \#2 and the other model phones the performance was worse. It is evident from the results that for these three phones the algorithm had difficulties producing an accurate estimate of the full route. Nonetheless, in all cases the accuracy is always markedly higher than that of the random case.

\begin{table}
	\centering
	\small
		\begin{tabular}{c|c|c|c|c|}
			\cline{2-5}
			& random & frequent & Alg.~\ref{alg:iterative-majority-vote} & combined \\
			\hline
			\multicolumn{1}{|c|}{Nexus 4 \#1} & 0.61 & 0.38 & 0.27 & 0.24 \\
			\hline
			\multicolumn{1}{|c|}{Nexus 4 \#2} & 0.63 & 0.61 & 0.59 & 0.52 \\
			\hline
			\multicolumn{1}{|c|}{Nexus 5} & 0.68 & 0.6 & 0.55 & 0.45 \\
			\hline
			\multicolumn{1}{|c|}{HTC Desire} & 0.65 & 0.59 & 0.5 & 0.45 \\
			\hline
		\end{tabular}
		\normalsize
	\caption{Levenshtein distance}
	\label{tab:LevenshteinDistance}
\end{table}

\begin{table}
	\centering
	\small
		\begin{tabular}{c|c|c|c|c|}
			\cline{2-5}
			& random & frequent & Alg.~\ref{alg:iterative-majority-vote} & combined \\
			\hline
			\multicolumn{1}{|c|}{Nexus 4 \#1} & 4\% & 38\% & 22\% & 45\% \\
			\hline
			\multicolumn{1}{|c|}{Nexus 4 \#2} & 5\% & 8.5\% & 5\% & 15\% \\
			\hline
			\multicolumn{1}{|c|}{Nexus 5} & 3\% & 15\% & 9\% & 20\% \\
			\hline
			\multicolumn{1}{|c|}{HTC Desire} & 5\% & 10\% & 12\% & 17\% \\
			\hline
		\end{tabular}
		\normalsize
	\caption{Exact full route fit}
	\label{tab:ExactFullRouteFit}
\end{table}

\begin{figure*}
  \centering
  \begin{subfigure}{0.3\textwidth}
    \includegraphics[width=\textwidth]{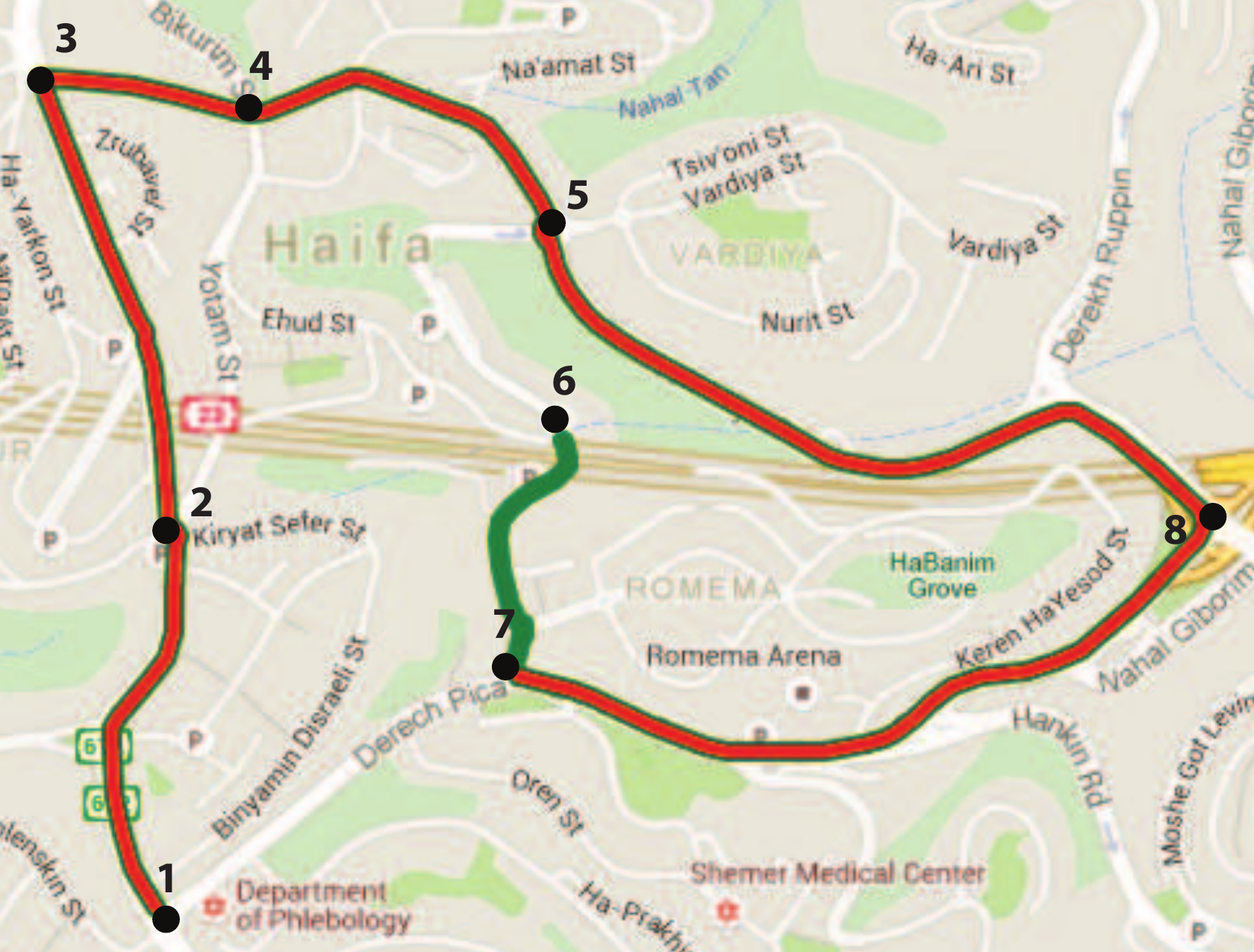}
    \caption{Distance = 0.125}
  \end{subfigure}
  \hspace{0.1cm}
  \begin{subfigure}{0.3\textwidth}
    \includegraphics[width=\textwidth]{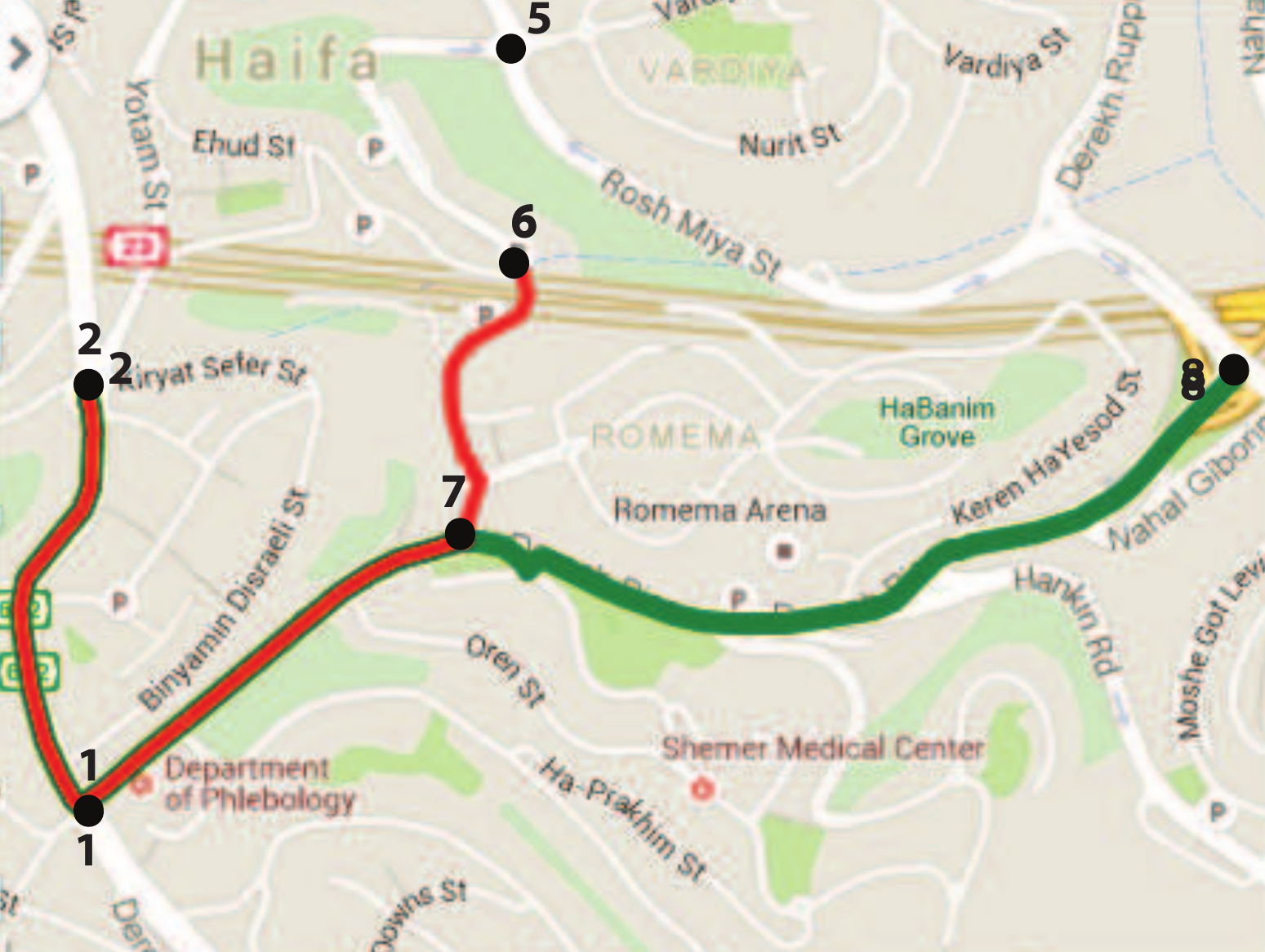}
    \caption{Distance = 0.25}
  \end{subfigure}
  \hspace{0.1cm}
  \begin{subfigure}{0.31\textwidth}
    \includegraphics[width=\textwidth]{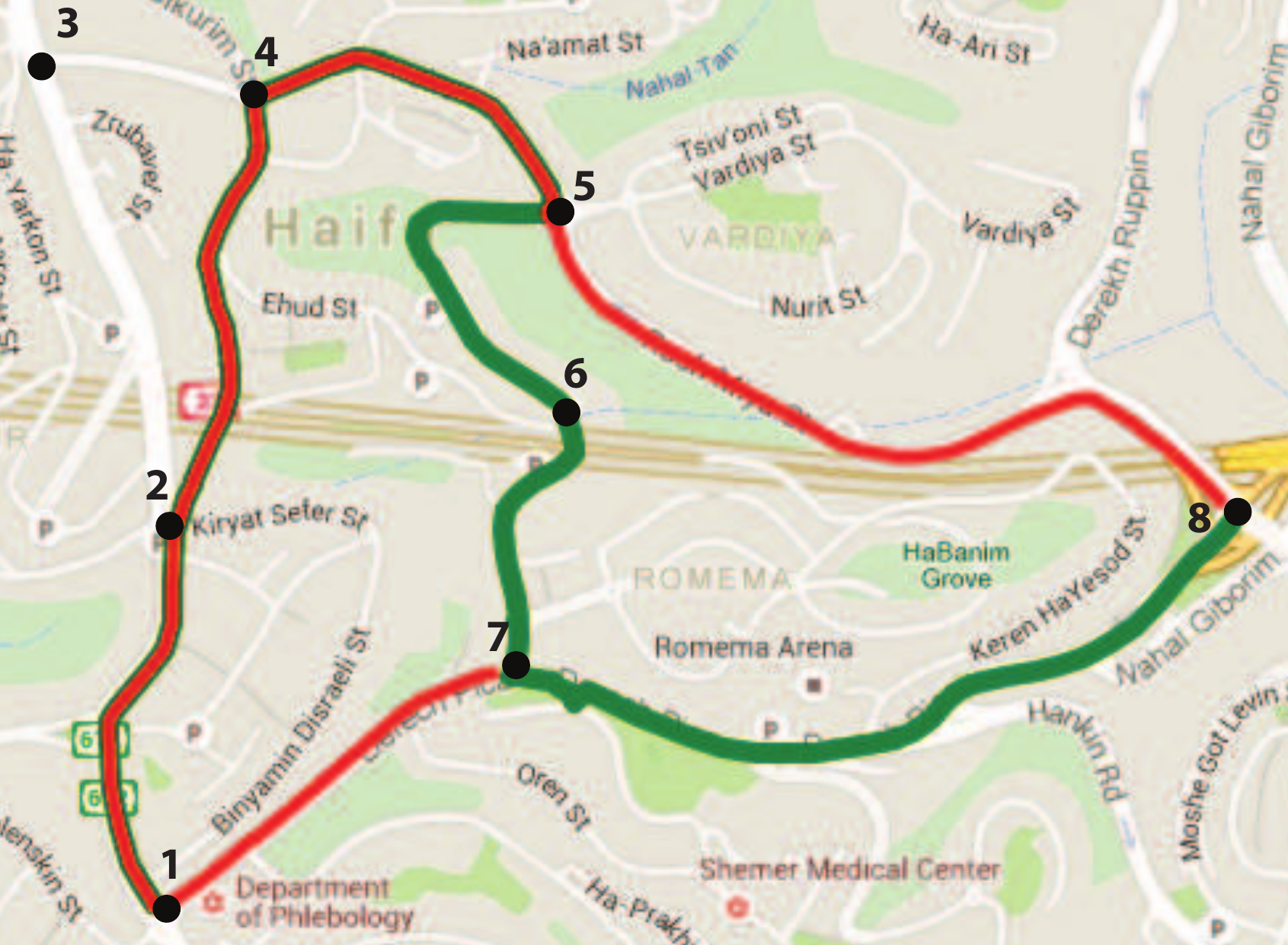}
    \caption{Distance = 0.43}
  \end{subfigure}
  \caption{Examples of estimation errors and their corresponding distances (partial map is depicted). The true route is green and the
  estimated route is red.}
  \label{fig:distance-error}
\end{figure*}

To have a better sense of the distance metric used to evaluate the quality of the estimated routes Figure~\ref{fig:distance-error} depicts three cases of estimation errors and their corresponding distance values in increasing order. It can be seen that even  estimation error having relatively high distances can have a significant amount of information regarding the true route.

\section{Future directions}  \label{sec:future}
In this section we discuss ideas for further research, improvements, and additions to our method.

\subsection{Power consumption inference}
While new (yet very common) smartphone models contain an internal ampere-meter and provide access
to current data, other models (for instance Galaxy S III) supply voltage but not current
measurements. Therefore on these models we cannot directly calculate the power consumption.
V-edge \cite{Xu2013} proposes using voltage dynamics to model a mobile device's power consumption.
That and any other similar technique would extend our method and make it applicable to additional
smartphone models.

Ref.~\cite{Zhang2010} presents PowerTutor, an application that estimates power consumption
by different components of the smartphone device based on voltage and state of discharge measurements.
Isolating the power consumed by the cellular connectivity will improve our method by eliminating
the noise introduced by other components such as audio/Bluetooth/WiFi etc. that do not directly depend
on the route.

\subsection{State of Discharge (SOD)}\todo{Is that too far fetched?}
The time derivative of the State-of-Discharge (the battery level) is basically a very coarse
indicator of power consumption. While it seemed to be too inaccurate for our purpose, there
is a chance that extracting better features from it or having few possible routes may
render distinguishing routes based on SOD profiles feasible. Putting it to the test is even more
interesting given the HTML 5 Battery API that enables obtaining certain battery statistics from a
web-page via JavaScript. Our findings demonstrate how future increases in the sampling
resolution of the battery stats may turn this API even more dangerous, allowing web-based attacks.

\subsection{Choice of reference routes}
Successful classification depends among other factors on good matching between the power profile we
want to classify and the reference power profiles.
Optimal matching might be a matter of month, time of day, traffic on the road, and more.
We can possibly improve our classification if we tag the reference profiles with those associated
conditions and select reference profiles matching the current conditions when trying to distinguish
a route. That of course requires collecting many more reference profiles.

\subsection{Collecting a massive dataset}
Collecting a massive dataset of power profiles associated with GPS coordinates is a feasible
task given vendors' capability to legally collect analytics about users' use of their smartphones.
Obtaining such big dataset will enable us to better understand how well our approach can scale and
whether it can be used with much less prior knowledge about the users.

\section{Defenses}
\label{sec:defenses}

\subsection{Non-defenses}
One might think that by adding noise or limiting the sampling rate or the resolution
of the voltage and current measurements one could protect location privacy.
However, our method does not rely on high sampling frequency or resolution.
In fact, our method works well with profiles much coarser than what we can directly
get from the raw power data, and for the route distinguishing task we actually
performed smoothing and downsampling of the data yet obtained good results.
Our method also works well with signal strength, which is provided with much lower
resolution and sampling frequency\footnote{In fact, since it
reflects more directly the environmental conditions, signal strength data can
provide even better route identification and tracking. We did not focus on signal strength
since accessing it requires access permissions and has already drawn research attention
to it as useful for localization.}.

\subsection{Risky combination of power data and network access}
One way of reporting voltage and current measurements to the attacker is via
a network connection to the attacker's server. Warning the user of this risky
combination may somewhat raise the bar for this attack.
There are of course other ways to leak this information. For instance, a malicious
application disguised as a diagnostic software can access power data and log
it to a file, without attempting to make a network connection,
while another, seemingly unrelated, application reads the data from that file
and sends it over the network.

\subsection{Secure hardware design}
The problem with access to total power consumption is that it leaks the
power consumed by the transceiver circuitry and communication related tasks
that indicate signal strength.
While power measurements can be useful for profiling applications, in many cases,
examining the power consumed by the processors executing the
software logic might be enough.
We therefore suggest that supplying only measurements of the power consumed by the processors
(excluding the power consumed by the TX/RX chain) could be a reasonable trade-off between
functionality and privacy.

\subsection{Requiring superuser privileges}
A simple yet effective prevention may be requiring superuser privileges
(or being root) to access power supply data on the phone. Thus, developers and
power-users can install diagnostic software or run a version of their application
that collects power data on a rooted phone, whereas the release version of the software
excludes this functionality.
This would of course prevent the collection of anonymous performance statistics from the
install-base, but as we have shown, such data can indicate much more than performance.

\subsection{Power consumption as a coarse location indicator}
Same as the cell identifier is defined as a coarse location indicator,
and requires appropriate permissions to be accessed, power consumption data
can also be defined as one.
The user will then be aware, when installing applications that access voltage
and current data, of the application's potential capabilities, and the
risk potentially posed to her privacy.
This defense may actually be the most consistent with the current security policies
of smartphone operating systems like Android and iOS, and their current permission
schemes.

\section{Related work}

Power analysis is known to be a powerful side-channel. The most well-known example is
the use of high sample rate ($\sim$20~MHz) power traces from externally connected
power monitors to recover private encryption keys from a cryptographic
system~\cite{kocher1999differential}.
Prior work has also established the relationship between signal strength and
power consumption in smartphones~\cite{Carroll2010, Schulman2010}. Further,
Bartendr~\cite{Schulman2010} demonstrated that paths of signal strength
measurements are stable across several drives.

PowerSpy combines these insights on power analysis and improving smartphone
energy efficiency to reveal a new privacy attack. Specifically, we demonstrate
that an attacker can determine a user's location simply by monitoring the
cellular modem's changes in power consumption with the smartphone's alarmingly
unprotected $\sim$100~Hz internal power monitor.

\subsection{Many sensors can leak location}

Prior work has demonstrated that data from cellular modems can be used to
localize a mobile device (an extensive overview appears in Gentile et
al.~\cite{Gentile2013}). Similar to PowerSpy, these works fingerprint the area
of interest with pre-recorded radio maps. Others use signal strength to
calculate distances to base stations at known locations.  All of these
methods~\cite{muthukrishnan2009inferring,krumm2004locadio,sohn2006mobility,ouyang2009received}
require signal strength measurements and base station ID or WiFi network name
(SSID), which is now protected on Android and iOS.  Our work does not rely on
the signal strength, cell ID, or SSID.  PowerSpy only requires access to power
measurements, which are currently unprotected on Android.

PowerSpy builds on a large body of work that has shown how a variety of unprotected sensors can leak location information.
Zhou et al. \cite{Zhou2013} reveal that audio on/off status is a side-channel
for location tracking without permissions. In particular, they extract a
sequence of intervals where audio is on and off while driving instructions
are being played by Google's navigation application. By comparing these intervals with
reference sequences, the authors were able to identify routes taken by the
user. \emph{SurroundSense}~\cite{azizyan2009surroundsense} demonstrates that
ambient sound and light can be used for mobile phone localization. They focus
on legitimate use-cases, but the same methods could be leveraged for breaching
privacy. \emph{ACComplice}~\cite{han2012accomplice} demonstrates how continuous
measurements from unprotected accelerometers in smartphones can reveal a user's
location. Hua et al.~\cite{hua2015metro} extend ACComplice by showing that
accelerometers can also reveal where a user is located in a metropolitan train
system.

\subsection{Other private information leaked from smartphone sensors}

An emerging line of work shows that various phone sensors can leak private
information other than location. In future work we will continue analyzing
power measurements to determine if other private information is leaked.

Prior work has demonstrated how smartphone sensors can be used to fingerprint
specific devices.
\emph{AccelPrint}~\cite{dey2014accelprint} shows that smartphones can be
fingerprinted by tracking imperfections in their accelerometer measurements.
Fingerprinting of mobile devices by the characteristics of their loudspeakers
is proposed in \cite{clarkson2012breaking,DASAN}. Further, Bojinov et. al.
~\cite{bojinov2014mobile} showed that various sensors in smartphones can be used
to identify a mobile device by its unique hardware characteristics.
Lukas et. al. \cite{lukas2006digital} proposed a method for digital camera
fingerprinting by noise patterns present in the images. \cite{li2010source}
enhances the method enabling identification of not only the model but also
particular cameras.

Sensors can also reveal a user's input such as speech and touch gestures.
The \emph{Gyrophone} study~\cite{michalevsky2014gyrophone} showed that
gyroscopes on smartphones can be used for eavesdropping on a conversation in
the vicinity of the phone and identifying the speakers.
Several works~\cite{aviv2012practicality, cai2011touchlogger, xu2012taplogger}
have shown that the accelerometer and gyroscope can leak information about
touch and swipe inputs to a foreground application.

\section{Conclusion}
PowerSpy shows that applications with access to a smartphone's power monitor can
gain information about the location of a mobile device -- without
accessing the GPS or any other coarse location indicators. Our
approach enables known route identification, real-time tracking, and
identification of a new route by only analyzing the phone's power
consumption.  We evaluated PowerSpy on real-world data
collected from popular smartphones that have a significant mobile
market share, and demonstrated their effectiveness.  We believe that
with more data, our approach can be made more accurate and reveal more
information about the phone's location.

Our work is an example of the unintended consequences that result from
giving 3rd party applications access to sensors.  It suggests that
even seemingly benign sensors need to be protected by permissions, or
at the very least, that more security modeling needs to be done before
giving 3rd party applications access to sensors.

\section*{Acknowledgments}

We would like to thank Gil Shotan and Yoav Shechtman for helping to collect
the data used for evaluation, Prof. Mykel J. Kochenderfer from Stanford University for providing
advice regarding location tracking techniques, Roy Frostig for providing advice regarding
classification and inference on graphs, and finally Katharina Roesler for proofreading the paper.
This work was supported by NSF and the DARPA SAFER program. Any opinions, findings and conclusions or 
recommendations expressed in this material are those of the author(s) and do not necessarily reflect the views of NSF or DARPA.

% trigger a \newpage just before the given reference
% number - used to balance the columns on the last page
% adjust value as needed - may need to be readjusted if
% the document is modified later
% \IEEEtriggeratref{16}
% The "triggered" command can be changed if desired:
%\IEEEtriggercmd{\enlargethispage{-5in}}

% references section

% can use a bibliography generated by BibTeX as a .bbl file
% BibTeX documentation can be easily obtained at:
% http://www.ctan.org/tex-archive/biblio/bibtex/contrib/doc/
% The IEEEtran BibTeX style support page is at:
% http://www.michaelshell.org/tex/ieeetran/bibtex/
%\bibliographystyle{IEEEtran}
% argument is your BibTeX string definitions and bibliography database(s)
%\bibliography{IEEEabrv,../bib/paper}
%
% <OR> manually copy in the resultant .bbl file
% set second argument of \begin to the number of references
% (used to reserve space for the reference number labels box)
{\footnotesize \bibliographystyle{acm}
\bibliography{powerspy}}

% \theendnotes

\appendix
\section{Formal model of new route inference} \label{sec:model}
In this section we formalize the problem of the new route inference (Section~\ref{sec:newroutes}) as a hidden Markov model (HMM) \cite{Rabiner1989}.
Let $I$ denote the set of intersections in an area in which we wish to track a mobile device.
A road segment is given by an ordered pair of intersections $(x, y)$, defined to be a continuous road between
intersection $x$ and intersection $y$.
We denote the set of road segments as $R$.

We assume that once a device starts to traverse a road segment it does not change the direction of
its movement until it reaches the end of the segment.
We define a state for each road segment. We say that the tracked device is in state $s_{xy}$ if the device is
currently traversing a road segment $(x, y)$, where $x,y \in I$.
We denote the route of the tracked device as a $(Q,T)$, where
\begin{gather*}
Q = \left\{q_{1}=s_{x_{1}x_{2}}, q_{2}=s_{x_{2}x_{3}}, ...\right\} \hspace{10px}
T = \left\{t_{1}, t_{2}, ...\right\}
\end{gather*}

For such a route the device has traversed from $x_{i}$ to $x_{i+1}$ during time interval
$[t_{i-1},t_i]$ ($t_0 = 0, t_{i-1} < t_i ~ \forall i > 0$).

Let $A=\left\{a_{xyz} | \forall x,y,z \in I\right\}$ be the state transition probability distribution,
where
\begin{equation}
a_{xyz} = p\left\{q_{i+1} = s_{yz} | q_{i} = s_{xy} \right\}
\end{equation}
Note that $a_{xyz} = 0$ if there is no road between intersections $x$ and $y$ or no road between intersections $y$ and $z$.
%For simplicity we shall assume a uniform probability distribution between the possible state transitions.
A traversal of the device over a road segment yields a power consumption profile of length equal to the duration of that movement.
%We define for each state $s_{xy}$ a normal time $T_{ij}$ which shall be the duration of a normalized power consumption profile for state $s_{ij}$.
We denote a power consumption profile as an observation $o$. %We denote $|o|$ as the duration of $o$.
Let $B$ be the probability distribution of yielding a given power profile while the device
traversed a given segment. Due to the hysteresis of hand-offs between cellular base stations, this probability depends on the previous segment the device traversed.
Finally, let $\Pi=\left\{\pi_{xy}\right\}$ be the initial state distribution, where $\pi_{xy}$ is the probability that the device initially traversed segment $(x,y)$. If there is no road segment between intersections $x$ and $y$, then $\pi_{xy}=0$. In our model we treat this initial state as the state of the device \emph{before} the start of the observed power profile. We need to take this state into account due to the hysteresis effect.  Note that an HMM is characterized by $A$, $B$, and $\Pi$.

The route inference problem is defined as follows.
Given an observation of a power profile $O$ over time interval $[0,t_{\max}]$,
and given a model $A$, $B$ and $\Pi$, we need to find a route $(Q,T)$ such that
$p\left\{(Q, T) | O\right\}$ is maximized.
In the following we denote the part of $O$ which begins at
time $t'$ and ends at time $t''$ by $O_{[t', t'']}$.
Note that $O=O_{[0, t_{\max}]}$.
We consider the time interval $[0,t_{\max}]$ as having a discrete resolution of $\tau$.
%We also assume there is a minimum time duration $\Delta^{xy}_{\min}$ for which a device can
%traverse a road segment $(x,y)$ (based on the device's maximum velocity and length of the road segment).

\section{Choosing the best inferred route} \label{sec:bestroute}
Upon its completion, the particle filter described in \cref{sec:particle-filter} outputs a set of $N$ routes of various lengths. We denote this set by $P_{\textrm{final}}$. 
This set exhibits an estimate of the distribution of routes given the power profile of the tracked device. 
The simple approach to select the best estimate is to choose the route that appears most frequently in $P_{\textrm{final}}$ as it has the highest probability to occur. Nonetheless, since a route is composed of multiple segments chosen at separate steps, at each step the weight of a route is determined solely based on the last segment added to the route. Therefore, in $P_{\textrm{final}}$ there is a bias in favor of routes ending with segments that were given higher weights, while the weights of the initial segments have a diminishing effect on the route distribution with every new iteration.

To counter this bias, we choose another estimate using a procedure we call \emph{iterative majority vote}. This procedure ranks the routes based on the prevalence of their prefixes. At each iteration $i$ the procedure calculates -- Prefix[i] -- a list of prefixes of length $i$ ranked by their prevalence out of the all routes that has a prefix in Prefix[i-1]. Prefix[i][n] denotes the prefix of rank $n$. The operation $p || j$ -- where $p$ is a route and $j$ is an intersection -- denotes the appending of $j$ to $p$. At each iteration $i$ \cref{alg:iterative-majority-vote} is executed. In the following we denote RoutePrefixed(R, p) to be the subset of routes out of the set $R$ having $p$ as their prefix.

\begin{algorithm}
    \begin{algorithmic}
    	\State $I' \gets I$
			\While{not all prefixes found}
				\State Prf $\gets$ next prefix from \textrm{Prefix}[i].
				\State Find $j \in I'$  that maximizes \\~~~~~~~~~$\textrm{RoutePrefixed}(\textrm{RoutePrefixed}(P_{\textrm{final}}, \textrm{Prf}), \textrm{Prf} || j)$
				\If{no such $j$ is found}
					\State $I' = I$
					\State continue loop
				\EndIf
				\State $\textrm{Prefix}[i+1] \gets \textrm{Prefix}[i+1] \cup \left\{\textrm{Prf}|| j\right\}$
				\State $I' = I' - \left\{j\right\}$
			\EndWhile
    \end{algorithmic}

    \caption{Iterative majority vote}
    \label{alg:iterative-majority-vote}
\end{algorithm}

At each iteration $i$ we rank the prefixes based on the ranks of the previous iteration. Namely, prefixes which are extensions of a shorter prefix having a higher rank in a previous iteration will always get higher ranking over prefixes which are extensions of a lower rank prefix. At each iteration the we first find the most common prefixes of length $i+1$, which start with the most common prefix of length $i$ found in the previous iteration, and rank them according to their prevalence. Then we look for common prefixes of length $i+1$, that start with the second most common prefix of length $i$ found in the previous iteration, and so on until all prefixes of length $i+1$ are found. The intuition is as follows. The procedure prefers routes traversing segments that are commonly traversed by other routes. Those received a high score when were chosen. Since we cannot pick the most common segments separately from each step (a continuous route probably will not emerge), we iteratively pick the most common segment out of the routes that are prefixed with the segments that were already chosen.

% that's all folks
\end{document}